\documentclass[aps,a4,twocolumn,superscriptaddress,preprintnumbers,nofootinbib]{revtex4-1}
\usepackage[usenames,dvipsnames]{color}  
\usepackage{graphicx}
\usepackage{pifont}
\usepackage{caption}
\usepackage{subcaption}
\usepackage{amsmath}
\usepackage{amssymb}
\usepackage[colorlinks=true,citecolor=darkred,urlcolor=darkred,pdfborder={0 0 0}]{hyperref}
\usepackage[normalem]{ulem}

\usepackage[T1]{fontenc}
\usepackage[latin9]{inputenc}
\usepackage{array}
\usepackage{booktabs}
\usepackage{mathrsfs}
\usepackage{multirow}
\usepackage{tabularx}
\usepackage[export]{adjustbox}
\usepackage{float}
\usepackage{multirow}

\definecolor{darkred}{rgb}{0.6,0,0}
\definecolor{dbrown}{rgb}{0.4,0.26,0.13}
\definecolor{linkcolor}{rgb}{0,0,0.5}

\definecolor{vdrgreen}{rgb}{0.0, 0.7, 0.0}

\begin{document}

\title{Impact of COHERENT measurements, cross section uncertainties\\
  and new interactions on the neutrino floor}%
\author{D. Aristizabal Sierra}%
\email{daristizabal@ulg.ac.be}%
\affiliation{Universidad T\'ecnica
  Federico Santa Mar\'{i}a - Departamento de F\'{i}sica\\
  Casilla 110-V, Avda. Espa\~na 1680, Valpara\'{i}so, Chile}%
\author{V. De Romeri}%
\email{deromeri@ific.uv.es}%
\affiliation{Instituto de F\'{i}sica Corpuscular,
  CSIC/Universitat de Val\`encia,\\ Calle Catedr\'atico Jos\'e
  Beltr\'an, 2 E-46980 Paterna, Spain}%
\author{L. J. Flores}%
\email{ljflores@jerez.tecnm.mx}%
\affiliation{Instituto de F\'isica, Universidad Nacional Aut\'onoma de M\'exico, A.P. 20-364, Ciudad de M\'exico 01000, M\'exico.}%
\affiliation{Tecnol\'ogico Nacional de M\'exico/ITS de Jerez, C.P. 99863, Zacatecas, M\'exico.}%
\author{D. K. Papoulias}%
\email{d.papoulias@uoi.gr}%
\affiliation{Department of Physics, University of Ioannina GR-45110
  Ioannina, Greece}%
\begin{abstract}
  We reconsider the discovery limit of multi-ton direct detection dark
  matter experiments in the light of recent measurements of the
  coherent elastic neutrino-nucleus scattering
  process. Assuming the cross section to be a parameter entirely
  determined by data, rather than using its Standard Model
  prediction, we use the COHERENT CsI and LAr data sets to determine
  WIMP discovery limits. Being based on a data-driven approach, the
  results are thus free from theoretical assumptions and fall within
  the WIMP mass regions where XENONnT and DARWIN have best expected
  sensitivities. We further determine the impact of subleading nuclear form
  factor and weak mixing angle uncertainties effects on WIMP discovery
  limits. We point out that these effects, albeit small, should be
  taken into account. Moreover, to quantify the impact of new physics effects in
  the neutrino background, we revisit WIMP discovery limits assuming
  light vector and scalar mediators as well as neutrino magnetic
  moments/transitions. We stress that the presence of new interactions
  in the neutrino sector, in general, tend to worsen the WIMP
  discovery limit.
\end{abstract}
\maketitle
\section{Introduction}
\label{sec:intro}
Cosmological and astrophysical data support the idea that dark matter
(DM) is the dominant form of matter in the Universe. One of the most
considered hypothesis is that of DM being a thermal species weakly
coupled to the thermal bath and whose abundance is determined by
thermal freeze-out (a species usually referred to as WIMP). The main
motivation for such a scenario is---arguably---the fact that its
abundance is entirely determined by the Universe expansion rate and
by interactions of DM with the early Universe thermal bath. This
means that once a cosmological and a particle physics model are
specified, the determination of the DM abundance is to a large extent
reduced to a parameter space-related question.  A rather large list of
such models exist and have been the subject of a great deal of
phenomenological and experimental activity, which includes---among
others---the direct detection of DM in laboratory experiments.

The DM direct detection program dates back to the early nineties, with the
first germanium ionization detectors using few kilogram target
material \cite{Baudis:2012ig}. The most up-to-date data, which lead to
the most stringent limits on the DM-nucleon cross section, follow from
measurements of order ton-size liquid xenon (LXe) time projection chambers (TPCs)
and include the LUX, PandaX-II and XENON1T experiments
\cite{LUX:2015abn,PandaX-II:2017hlx,XENON:2019gfn}. Measurements on
liquid argon (LAr) TPCs, which include DarkSide-50 and DEAP-3600, have as well placed
limits, albeit less stringent due to their lower exposures and higher
recoil energy thresholds \cite{DarkSide:2018kuk,DEAP:2019yzn}. In the
next few years searches will continue, with LXe TPC experiments paving
the way. Future experiments include LZ, XENONnT and ultimately DARWIN,
detectors which involve multi-ton fiducial volumes
\cite{Akerib:2018lyp,Aprile:2015uzo,Aalbers:2016jon,XENON:2020kmp}. The advent of
the multi-ton era implies that DM searches will be subject to
irreducible neutrino backgrounds, in particular those emitted in the
$^8$B process of the solar pp chain
\cite{Strigari:2009bq,Billard:2013qya}.

Neutrino backgrounds induce coherent elastic-neutrino nucleus
scattering (CE$\nu$NS) and so produce nuclear recoil spectra, which,
depending on the WIMP parameter space, can have a strong degeneracy
with those expected from spin-independent WIMP interactions\footnote{Certain spin dependent or spin and velocity dependent WIMP
  interactions can also induce recoil spectra that degenerate with the
  neutrino recoil spectra \cite{Dent:2016iht}.}. Actually, a full
degeneracy is found between $^8$B solar (atmospheric) neutrinos and a
WIMP model defined by a WIMP mass $m_\chi\simeq 6\,$GeV and
a WIMP-nucleon cross section $\sigma_{n-\chi}\simeq 5\times 10^{-45}\,\text{cm}^2$
($m_\chi\simeq 100\,$GeV and
$\sigma_{n-\chi}\simeq 10^{-48}\,\text{cm}^2$)
\cite{Billard:2013qya}. This level of degeneracy thus leads to a
saturation of the WIMP-nucleon cross section to which a particular experiment
can have access. So, in contrast to the background-free paradigm,
increasing exposure does not imply a linear improvement of
sensitivities but rather a saturation of its discovery limit
\cite{Strigari:2009bq}, typically referred to as \textit{neutrino
  floor}. Various experimental techniques that enable overcoming the
neutrino floor have been discussed in the literature. They include
measurements of the WIMP and neutrino recoil spectra tails
\cite{Ruppin:2014bra}, directionality (see
e.g. \cite{Vahsen:2021gnb}), measurements with different material
targets \cite{Ruppin:2014bra} and annual modulation
\cite{Davis:2014ama}. However, although feasible in principle, some of
them require large exposures and/or further technological improvements.

The experimental reach of multi-ton DM direct detection experiments
(with no directional capabilities) thus depends crucially on the
precision with which WIMP and CE$\nu$NS induced events can be
predicted. WIMP event rates are subject to astrophysical
uncertainties, which depend e.g. on the DM halo model assumed for
their calculation. Their impact have been studied in detail in
Ref.~\cite{OHare:2016pjy}. CE$\nu$NS event rate uncertainties instead
can be thought as being of two types, those associated with neutrino
flux normalizations and those associated with the CE$\nu$NS cross
section. The Standard Model (SM) CE$\nu$NS cross section uncertainties are mainly driven by
nuclear physics effects, encoded in the weak-charge form factor
\cite{AristizabalSierra:2019zmy,Papoulias:2019lfi,Hoferichter:2020osn}. 
For solar neutrinos these effects barely exceed $\sim 1\%$, while for
atmospheric neutrinos they can be larger but never exceeding
$\sim 10\%$.  For this reason, the neutrino flux normalization uncertainties
dominate the determination of the experimental reach a given
experiment can have.

The advent of the multi-ton era requires an
understanding of the discovery reach beyond that implied by the
neutrino flux normalization factors uncertainties. Since the effects
of astrophysical uncertainties have been already quantified, and have
been proved to have a small effect~\cite{OHare:2016pjy}, for this task
one should rather focus on the uncertainties in the neutrino
sector. In order to do so one can adopt a data-driven approach or
instead consider all possible effects that might have an impact on the
discovery potential. This paper aims at exploring both cases for LXe and LAr
detectors. With \textit{data-driven} analysis we mean using COHERENT data
\cite{Akimov:2017ade,Akimov:2018vzs,COHERENT:2020iec} to extract the
CE$\nu$NS cross section along with its uncertainty. The advantage of
this approach is that in such a way the cross section uncertainty encapsulates all possible effects, including
possible new physics contributions, without the need of any further
assumption. We present as well a more assumption-dependent analysis in
which we consider what could be regarded as sub-leading
uncertainties. These include effects related with possible low-energy
variations of the weak mixing angle and the unknown value of the xenon
point-neutron distribution mean-square radius. Finally, given the precision
with which CE$\nu$NS has been currently measured, possible new physics effects
can have a big impact too (see for instance \cite{Dutta:2017nht,Bertuzzo:2017tuf,AristizabalSierra:2017joc,Boehm:2018sux,Essig:2018tss,Denton:2018xmq,AristizabalSierra:2019ykk}).
 Here, we present an analysis of such effects by
considering new vector and scalar interactions in the light regime as
well as neutrino magnetic moments/transitions. 

The remainder of this paper is organized as follows. In
Sec. \ref{sec:method} we discuss WIMP and neutrino event rate spectra along
with the likelihood method that we use for the determination of discovery
limits. In Sec. \ref{sec:new-physics} we present the results of our
analyses obtained following the data-driven approach and considering
sub-leading uncertainty effects and new interactions. In
Sec. \ref{sec:conclusions} we present our conclusions. Finally, in Appendix
\ref{sec:extraction} we provide details of the procedure used for the
extraction of the CE$\nu$NS cross section from COHERENT data.
\section{WIMP and neutrino event rates}
\label{sec:method}
In this section we present a brief discussion of the event rates
induced by the interactions of the DM particles (in the local DM halo)
and neutrinos with the nuclear target material of a generic
detector. On dimensional grounds, event rates can be estimated to be
given by the number of scatterers $N_N$, the incident particle flux
$\Phi$ and the interaction probability of the incident particles with
the scatterers $\sigma$. In terms of these variables, the number of
expected events per unit of amount of target material and per time is
given by $R\sim N_N\times \Phi \times \sigma$. For DM, astrophysical
assumptions on the DM halo model are required to predict the WIMP flux
at the detector. As for neutrinos, fluxes fall in three categories:
solar, diffuse supernova neutrino background (DSNB) and sub-GeV atmospheric
neutrinos. Whether a certain type or component matters or not for a
certain detector depends on the energy threshold. In particular, for LXe 
detectors it is known that only the $^8$B component of the solar
neutrino spectrum matters \cite{Billard:2013qya,Ruppin:2014bra}. 
Here however we consider all components, which allows to extend our
analysis to a wider DM mass range.

Discussion of solar, DSNB and sub-GeV atmospheric neutrino fluxes have
been presented in a wide spectrum of references. As backgrounds for DM
direct detection searches they have been discussed in detail in
Refs. \cite{Monroe:2007xp,Vergados:2008jp,Strigari:2009bq,Gutlein:2010tq,Billard:2013qya,OHare:2016pjy,Gonzalez-Garcia:2018dep,AristizabalSierra:2017joc}.
Some of their main properties can be found also in standard textbooks
(see e.g. \cite{Giunti:2007ry}). Briefly, solar neutrinos are
generated in certain sub-processes of the pp chain cycle which
accounts for the hydrogen-helium fusion process responsible for most
of the solar energy. They are produced as well in the
carbon-nitrogen-oxygen (CNO) cycle, which for a main sequence star
accounts for less then $\sim 2\%$ of its energy. The neutrino spectrum
from the pp chain involves three monochromatic lines at
$E_\nu=0.38\,$MeV ($^7\text{Be}+e^-\to \nu_e+ ^7\text{Li}^*$),
$E_\nu=0.86\,$MeV ($^7\text{Be}+e^-\to \nu_e+ ^7\text{Li}$) and
$E_\nu=1.4\,$MeV (pep), along with three continuous spectra (pp, $^8$B
and hep) which extend up to energies of order $16\,$MeV (hep). The CNO
cycle involves instead three continuous spectra ($^{13}$N, $^{15}$O,
$^{17}$F) whose kinematic tails are located at $\sim 1.2\,$MeV.

The DSNB stems from the cumulative flux of neutrinos from supernova
(SN) explosions all over the history of the Universe. Compared to
solar neutrino fluxes it is less abundant but it matters once the hep
neutrino flux reaches its kinematic tail. Since DSNB neutrino energies
are determined by SN dynamics, the flux extends only up to
$E_\nu\sim 50\,$MeV. At $E_\nu\sim 30\,$MeV however, the sub-GeV
atmospheric neutrino flux kicks in and dominates the neutrino spectrum
up to the energies that matter for CE$\nu$NS,
$\sim 200-300\,$MeV. Atmospheric neutrinos arise from cosmic rays
interactions with the Earth atmosphere and the subsequent decays of
pions and muons.

For solar neutrinos, predictions from the B16-GS98 high-metallicity
standard solar model (SSM) \cite{Vinyoles:2016djt} are used. For DSNB neutrino fluxes we
instead use values that follow from theoretical predictions relying
on the assumption that the SN neutrino spectrum is well
approximated by a Fermi distribution with temperatures in the 3-8 MeV
range \cite{Beacom:2010kk,Strigari:2009bq}. For sub-GeV atmospheric
neutrinos we use the predictions obtained in Ref. \cite{Honda:2015fha}
from Monte Carlo simulations of cosmic-ray cascades. Note that in
contrast to WIMP fluxes as well as DSNB and sub-GeV atmospheric
neutrino fluxes, almost all solar neutrino flux components have been
measured, with neutrinos from the $^{15}$O CNO cycle subprocess being
the most recent measurement
\cite{SNO:2011hxd,Borexino:2017rsf,SNO:2006dke,BOREXINO:2020aww}. 
Since DM direct detection experiments rely on nuclear recoil
measurements, solar neutrino events are dominated by the $^8$B
neutrino flux. Measurements of this flux have been performed at
Super-Kamiokande (SK) and BOREXINO using neutrino-electron elastic
scattering events and at SNO using neutrino scattering on deuteron
\cite{Super-Kamiokande:2005wtt,Borexino:2008fkj,SNO:2009uok}. Exposures
at BOREXINO are of the order of 300 tonne-year, while at SK and SNO
above 1000 tonne-year. These numbers imply that DM detectors could
provide complementary information on the $^8$B neutrino flux (nuclear
channel instead of electron channel), but will not have the capability
to improve upon the uncertainties these experiments have placed. In
contrast, in the atmospheric sector they can provide the first ever
measurement of sub-GeV neutrino fluxes. This will require exposures of
the order of 700 tonne-year, but provided they can be achieved this
will lead to a $5\sigma$ observation \cite{Newstead:2020fie}. Direct
measurement of the atmospheric component will certainly reduce current
uncertainties, entirely determined by Monte Carlo simulations.
Neutrino
flux normalization factors along with their uncertainties are
displayed in Tab. \ref{tab:parameters}.

\begin{table*}
  \renewcommand{\tabcolsep}{0.35cm}
  \renewcommand{\arraystretch}{1.4}
  \centering
  \begin{tabular}{|c|c|c||c|c|c|}\hline
    \multicolumn{6}{|c|}{\textbf{Neutrino flux components normalizations and uncertainties}}\\\hline
    Comp. & Norm. [$\text{cm}^{-2}\cdot\text{s}^{-1}$] & Unc.  & 
    Comp. & Norm. [$\text{cm}^{-2}\cdot\text{s}^{-1}$] & Unc.  \\\hline
    $^7$Be (0.38 MeV) & $4.84\times 10^8$&3\%&
    $^7$Be (0.86 MeV) & $4.35\times 10^9$&3\%\\\hline
    pep & $1.44\times 10^8$&1\% &
    pp & $5.98\times 10^{10}$&0.6\%\\\hline
    $^8$B & $5.25\times 10^6$&4\%&
    hep & $7.98\times 10^3$&30\%\\\hline
    $^{13}$N & $2.78\times 10^8$&15\%&
    $^{15}$O & $2.05\times 10^8$&17\%\\\hline
    $^{17}$F & $5.29\times 10^6$&20\%&
    DSNB & 86 &50\%\\\hline
    Atm & 10.5 &20\%&---&
    ---&---\\\hline
  \end{tabular}
  \caption{Neutrino flux normalization factors along with their 
    uncertainties as predicted by the B16-GS98 high metallicity SSM
    \cite{Vinyoles:2016djt}. Values follow the recommendations
    pointed out in Ref. \cite{Baxter:2021pqo}. These values along 
    with those in Tab. \ref{tab:parameters-WIMP} are used in the
    determination of WIMP discovery limits.}
  \label{tab:parameters}
\end{table*}

The CE$\nu$NS differential recoil spectrum follows from a convolution
of neutrino fluxes and the CE$\nu$NS differential cross section. For
the $\alpha$-th flux component it reads
\begin{equation}
  \label{eq:DRS_CEvNS}
  \frac{dR_\nu}{dE_r} = \varepsilon\frac{N_A}{m_\text{target}}
  \int_{E_\nu^\text{min}}^{E_{\nu}^\text{max}}\,\frac{d\Phi_\alpha}{dE_\nu}
  \frac{d\sigma}{dE_r}dE_\nu\ .
\end{equation}
Here $\varepsilon$ refers to the exposure in ton$\cdot$year units, $N_A$ to the
Avogadro number in mol$^{-1}$ units, $m_\text{target}$ to the nuclear
target molar mass and $d\Phi_\alpha/dE_\nu$ to the neutrino flux
(including its normalization). The integration lower limit is
determined by the recoil energy according to
$E_\nu^\text{min}=\sqrt{m_N E_r/2}$, with $m_N$ the scatterers' nuclear
mass. Since xenon has 9 stable isotopes, of which few of them have
substantially large natural abundances, in our analyses we work with
averaged nuclear mass and mass number: $m_N=\sum_i m_i X_i$ and
$A=\sum_i A_i X_i$, with the sum running over all stable isotopes. The
integration upper limit is determined by the flux kinematic tail. The
total number of CE$\nu$NS events induced by the $\alpha$-th flux is thus given by
\begin{equation}
  \label{eq:N_nu}
  N_\nu(\Phi_\alpha) = \int_{E_r^\text{min}}^{E_r^\text{max}}
  \frac{dR_\nu}{dE_r}dE_r
\end{equation}
where $E_r^\text{min}$ refers to recoil energy threshold and
$E_r^\text{max}\simeq 2E_\nu^2|_\text{tail}/m_N$.

The CE$\nu$NS differential scattering cross section, determined by a
neutral current process, is given by
\cite{Freedman:1973yd,Freedman:1977xn}
\begin{equation}
  \label{eq:CEvNS-xsec}
  \frac{d\sigma}{dE_r} = \frac{m_NG_F}{2\pi}Q_W^2F^2_W(q)
  \left(2-\frac{m_NE_r}{E_\nu^2}\right)\ ,
\end{equation}
where $Q_W$ is the coherent weak charge that quantifies the coupling
of the $Z$ gauge boson to the nucleus. It is therefore entirely
determined by electroweak $Z-q$ couplings, namely
\begin{equation}
  \label{eq:EW-Coherent-charge}
  Q_W = (A-Z) (g_V^u + 2 g_V^d) + Z (2g_V^u + g_V^d)\ ,
\end{equation}
with the couplings given by $g_V^u=1/2 - 4/3\sin^2\theta_W$ and
$g_V^d=-1/2+2/3\sin^2\theta_W$. For the weak mixing angle, in our
analyses, we use its low-energy value obtained by the RGE extrapolation
from the $Z$ scale to $q^2=0$, $\sin^2\theta_W=0.2387$
\cite{Kumar:2013yoa}. The cross section in
Eq.(\ref{eq:CEvNS-xsec}) comes along with the
weak-charge nuclear form factor which, combined with the coherent weak
charge, determines the $q$-dependent strength of the $Z$-nucleus
coupling. Throughout the paper we use the Helm parametrization
\cite{Helm:1956zz}. For the impact of uncertainties due to the
variations of the point-neutron distribution mean-square-radius
($R_n$), however, we express the weak-charge form factor in terms of
the spin-independent proton and neutron form factors (for which, again, we use the Helm parametrization), neglecting nucleon form factor
$q$-dependent terms (more details are given in
Sec.~\ref{sec:impact-quenching-FF}). Note that choosing a particular
form factor parametrization implies that our results involve,
depending on the momentum transfer, up to a $\sim 10\%$ theoretical
uncertainty for both the WIMP and CE$\nu$NS event rates
\cite{AristizabalSierra:2021uob,daristi:2021}.

\begin{figure*}
  \centering
  \includegraphics[scale=0.45]{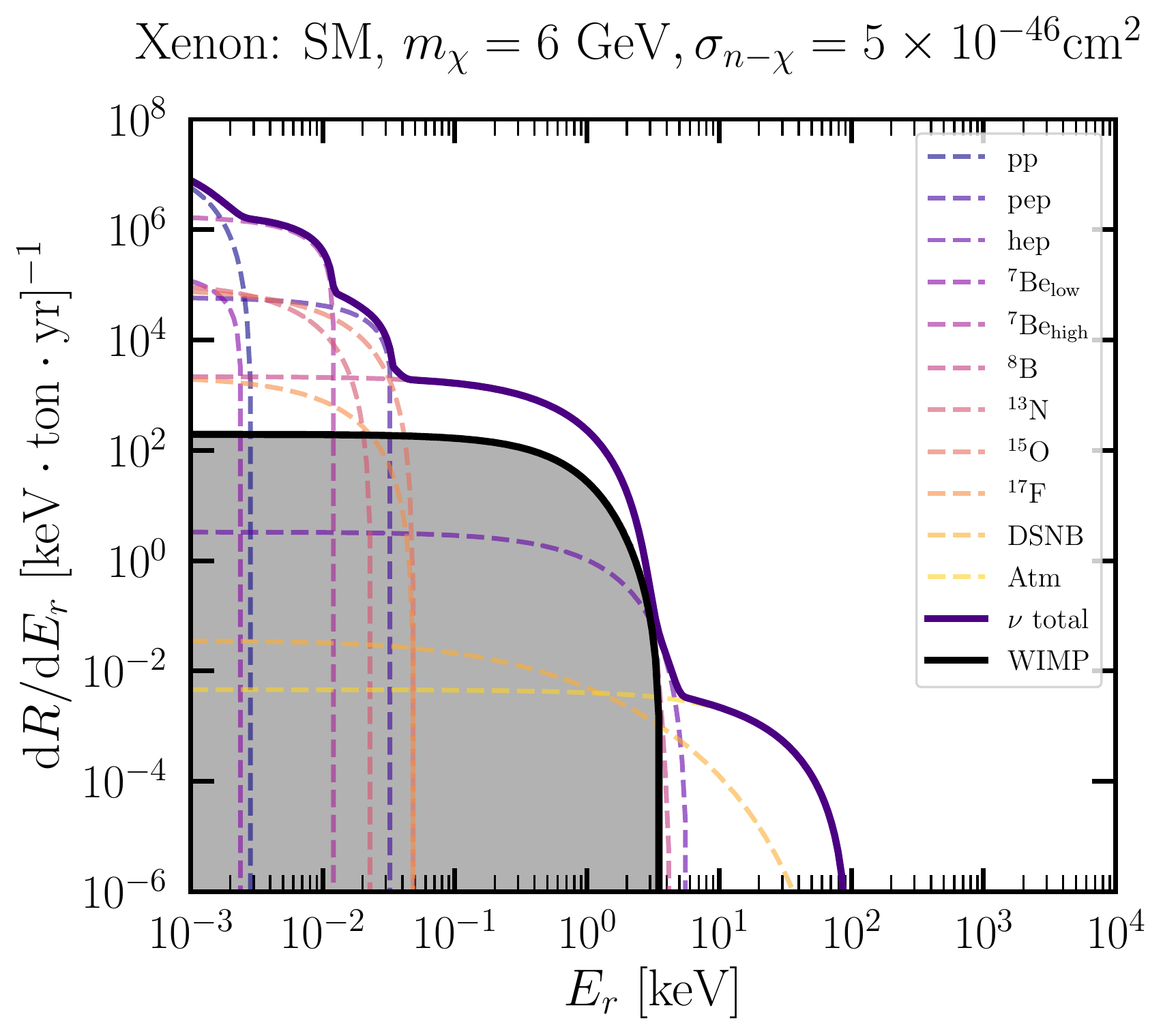}
  \hspace{0.5cm}
  \includegraphics[scale=0.45]{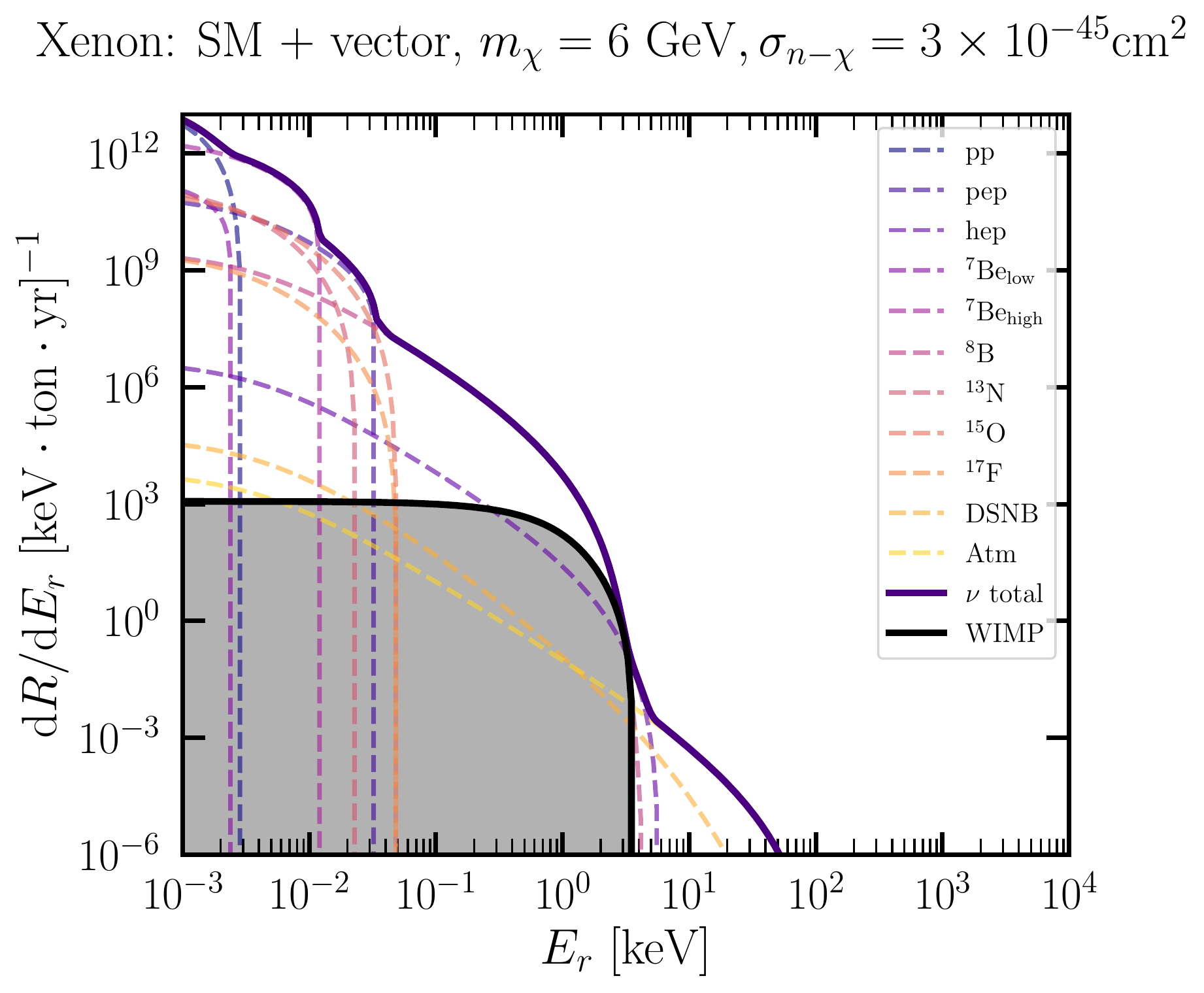}
  \caption{\textbf{Left graph}: Neutrino and WIMP differential recoil spectra expected in the SM. \textbf{Right graph}: Neutrino and WIMP differential recoil spectra in the presence of a long-range vector interaction. }
  \label{fig:diff_rates}
\end{figure*}

\begin{table*}
  \renewcommand{\tabcolsep}{0.35cm}
  \renewcommand{\arraystretch}{1.4}
  \centering
  \begin{tabular}{|c|c|c||c|}\hline
    \multicolumn{4}{|c|}{\textbf{Relevant WIMP related parameters}}\\\hline
    $v_0\,\text{[km/s]}$ & $v_\text{lab}\,\text{[km/s]}$
    & $v_\text{esc}\,\text{[km/s]}$ & $\rho_0\,\text{[GeV/cm$^3$]}$\\\hline
    220 & 232 & 544 & 0.3\\\hline
  \end{tabular}
  \caption{Values for the average, laboratory and escape velocities
    along with the local halo DM density $\rho_0=\rho(R_0=8\;\text{kpc})$
    used in the determination of WIMP discovery limits.}
  \label{tab:parameters-WIMP}
\end{table*}
On the other hand, the WIMP differential recoil spectrum can be
written according to
\begin{equation}
  \label{eq:wimp-DRS}
  \frac{dR_W}{dE_r} = \varepsilon\frac{\rho_0\sigma_\text{SI}(q)}{2m_\chi\mu^2}
  \int_{|\boldsymbol{v}|>v_\text{min}}d^3v\,\frac{f(\boldsymbol{v})}{v}\ ,
\end{equation}
where $\rho_0=\rho(R_0)$ ($R_0=8\,$kpc) is the local halo DM density,
$\sigma_\text{SI}(q)$ is the spin-independent
momentum-transfer-dependent WIMP-nucleus scattering cross section,
$m_\chi$ is the WIMP mass and $\mu$ is the WIMP-nucleus reduced
mass: $\mu = m_\chi m_N /(m_\chi + m_N)$. The integral corresponds to the mean inverse speed and its value
is determined by the assumed velocity distribution. The minimum
WIMP velocity, $v_\text{min}$, that can induce a nuclear recoil with
energy $E_r$ depends on whether the scattering is elastic
($\chi+N\to \chi+N$) or inelastic ($\chi+N\to \chi^\prime+N$). For
elastic scattering, for which our results apply\footnote{Inelastic contributions to the event rate are suppressed~\cite{Sahu:2020kwh}.}, one finds
$v_\text{min}=\sqrt{m_N E_r/2}/\mu$. 
As for illustration, we show in Fig.~\ref{fig:diff_rates} the neutrino and WIMP differential recoil spectra expected in the SM (left panel) and
in a new physics scenario with a light vector mediator (right panel), which we will discuss in more detail in section~\ref{sec:wimp-and-new-bckg}.
The WIMP mass has been fixed to 6 GeV and the WIMP-nucleon
momentum-transfer-independent cross section has been taken along the corresponding WIMP discovery limit shown in Fig.~\ref{fig:vector_scalar}
(left), obtained assuming a xenon detector with an exposure of 1 ton$\cdot$yr. This choice of WIMP mass and cross section leads to a WIMP differential rate which 
mimics almost exactly the differential rate of $^8$B solar neutrinos.

The total number of WIMP events
is obtained by integrating Eq.(\ref{eq:wimp-DRS})
\begin{equation}
  \label{eq:N_W}
  N_W=\int_{E_r^\text{min}}^{E_\text{max}}\frac{dR_W}{dE_r}dE_r\ ,
\end{equation}
where $E_\text{max}=2 \mu^2 (v_\text{esc}+v_\text{lab})^2/m_N$ (see
discussion below) \cite{Lewin:1995rx}.

In scenarios where the WIMP-proton and WIMP-neutron scattering cross
sections are equal (spin-conserving scenarios, $f_p/f_n=1$), and
nucleon form factor $q$-dependent terms are neglected,
$\sigma_\text{SI}(q)$ can be written as
\begin{equation}
  \label{eq:WIMP-N_WIMP-nucleon_xsec}
  \sigma_\text{SI}(q) = \frac{\mu^2}{\mu_n^2}
  \left[
    ZF_p(q) + (A-Z)F_n(q)
  \right]^2\sigma_{\chi-n}\ ,
\end{equation}
where $\sigma_{\chi-n}$ is the WIMP-nucleon
momentum-transfer-independent cross section and $\mu_n$  refers to the
WIMP-nucleon  reduced mass, $\mu_n = m_\chi m_n /(m_\chi + m_n)$ with $m_n=931.5$~MeV. This expression is particularly useful in
the treatment of uncertainties related with the point-neutron
distribution mean-square-radius. Here $F_{n,p}(q)$ are
spin-independent neutron and proton nuclear form factors
which, as in the neutrino sector, we parametrize \`a la Helm. Of
course if one assumes the point-nucleon distribution mean-square-radii
to be equal a much more simple (and familiar) relation follows
\begin{equation}
  \label{eq:WIMP-nucleus-WIMP-nucleon-xsecs}
  \sigma_\text{SI}(q) = \frac{\mu^2}{\mu^2_n} A^2 \sigma_{n-\chi}F^2(q^2)\ .
\end{equation}
In all our analyses, apart from that related with form factor
uncertainties, we make this simplifying assumption (see
Sec. \ref{sec:impact-quenching-FF} for further details).

The results presented in the following sections are obtained assuming
the standard halo model (SHM) \cite{Drukier:1986tm,Freese:1987wu},
which assumes that the local DM halo is dominated by a smooth and
virialized component (non-virialized components, such as streams or
debris flows, are regarded as subleading), well described by an
isothermal sphere with an isotropic and Maxwellian velocity
distribution according to
\begin{equation}
  \label{eq:Maxwell_velocity_dist}
  f(v)=
  \begin{cases}
    \frac{1}{N_\text{esc}}\left(\frac{3}{2\pi\sigma_v^2}\right)^{3/2}
    e^{-3\boldsymbol{v}^2/2\sigma_v^2}
    &\quad\text{for}\quad v<v_\text{esc}\ ,\\
    0&\quad\text{for}\quad v>v_\text{esc}\ ,
  \end{cases}\ .
\end{equation}
where $\sigma_v$ refers to the root-mean-square velocity dispersion
which determines the average (most likely) speed
$v_0=\sqrt{2/3}\sigma_v$. The normalization factor is in turn given by
\begin{equation}
  \label{eq:normalization_Max_distribution}
  N_\text{esc}=\text{erf}\left(\frac{v_\text{esc}}{v_0}\right)
  - \frac{2}{\sqrt{\pi}}\frac{v_\text{esc}}{v_0}e^{-v_\text{esc}^2/v_0^2}\ .
\end{equation}
The Maxwellian distribution is truncated at the escape velocity,
$v_\text{esc}$, to account for the fact that for larger values the DM
is not any more gravitationally bounded and thus can escape the Galaxy
gravitational pull. With $f(v)$ defined as in
Eq.(\ref{eq:Maxwell_velocity_dist}), and after a Galilean boost to
the laboratory (Earth) frame (with velocity $v_\text{lab}$), the mean
inverse speed can be analytically calculated. Being a standard
well-known result it can be found in many references. We point the
reader to e.g. Ref. \cite{Freese:2012xd}. Values for the relevant
parameters used in our calculation are shown in
Tab. \ref{tab:parameters-WIMP}.
\subsection{WIMP discovery limits: Statistical approach}
\label{sec:statisics}
In this section we describe the statistical procedure adopted for the
determination of WIMP discovery limits, which follows a frequentist
significance test using a likelihood ratio as a test statistic
\cite{Cowan:2010js}. As a tool for the determination of WIMP discovery
limits, this technique was first used in Ref. \cite{Billard:2013qya}
and subsequently in
Refs. \cite{Ruppin:2014bra,OHare:2015utx,OHare:2016pjy,Gonzalez-Garcia:2018dep,OHare:2020lva}. In
general, both the calculation of signal (WIMP) and background (CE$\nu$NS) events may involve
nuisance parameters. We consider them only in the latter, assuming that they originate from uncertainties on the normalization of neutrino fluxes alone
(Sec. \ref{sec:wimp-and-new-bckg}) or combined with: (i) measured CE$\nu$NS cross
section uncertainties (Sec. \ref{sec:data-driven}), (ii) point-neutron
distribution mean-square-radius uncertainties
(Sec. \ref{sec:impact-quenching-FF}), (iii) weak mixing angle
uncertainties (Sec. \ref{sec:impact-quenching-FF}).

\begin{figure*}
  \centering
  \includegraphics[scale=0.6]{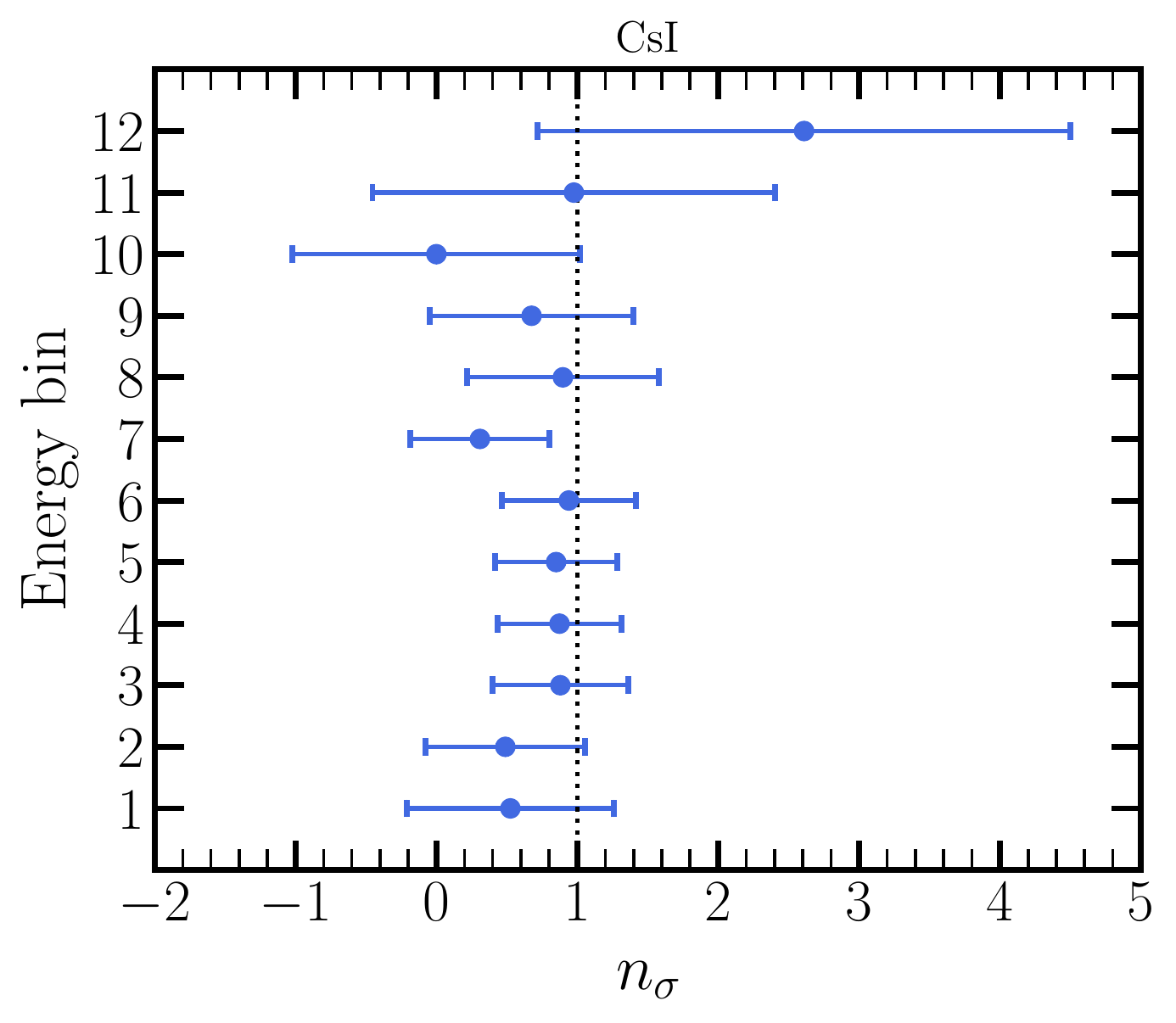}
  \hspace{0.5cm}
  \includegraphics[scale=0.6]{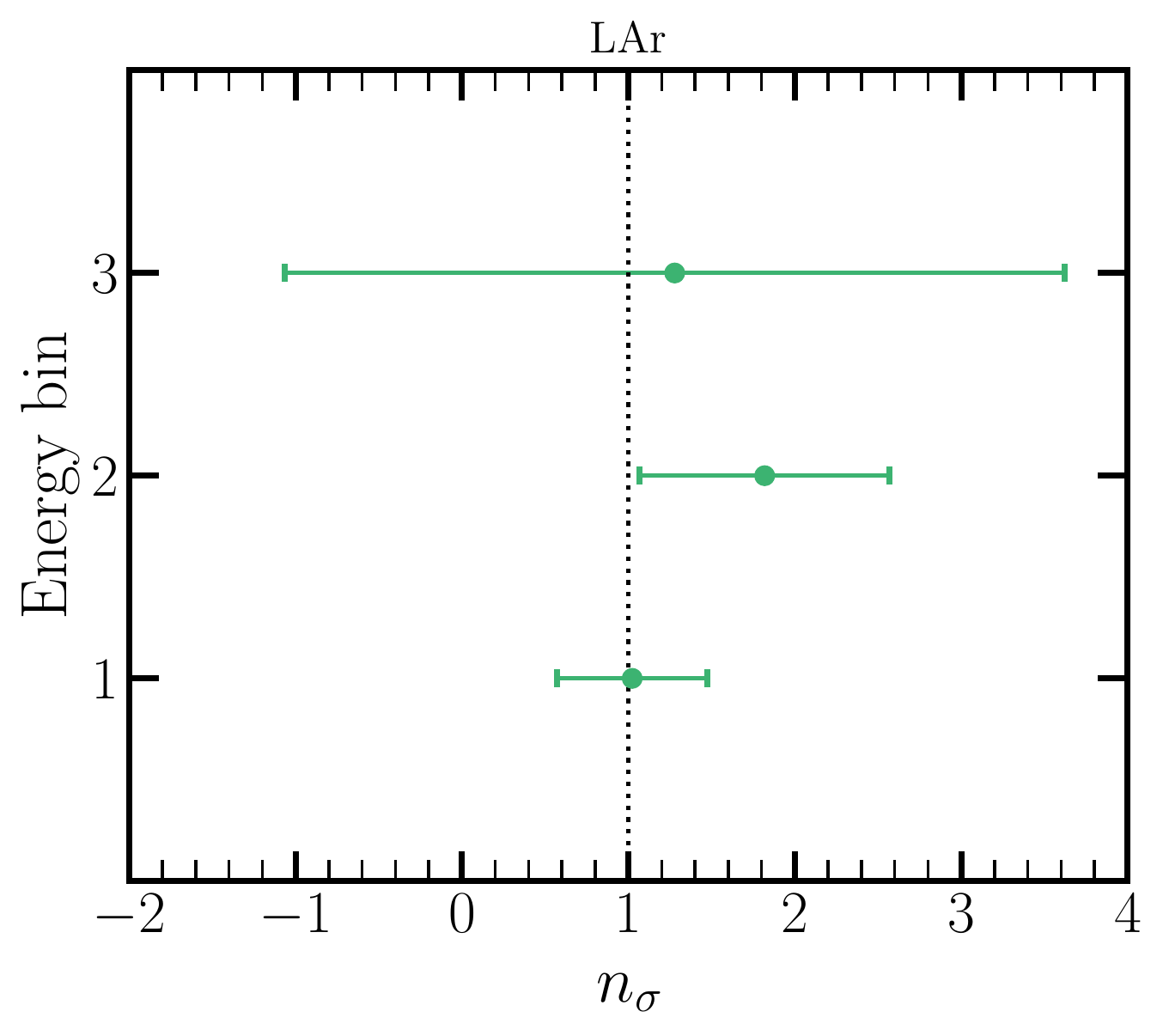}
  \caption{Experimentally measured CE$\nu$NS cross section normalized
    to the SM prediction ($n_\sigma$), extracted from COHERENT CsI and LAr data
    \cite{Akimov:2017ade,COHERENT:2020iec}. Results in each recoil energy bin
    indicate the central value (mean) alone with its uncertainty
    ($1\,\sigma$ CL). They are used in the data-driven analysis
    profile likelihood ratio test in Sec.~\ref{sec:data-driven}.}
  \label{fig:CEvNS_xsec_CsI_LAr}
\end{figure*}

The general likelihood function we adopt depends on WIMP parameters
($m_\chi$ and $\sigma_{n-\chi}$) as well as on the nuisance
parameters associated with neutrino fluxes normalization factors
(denoted $\phi_\alpha$, with $\alpha=1,\cdots , n_\nu=11$) and nuisance
$\mathcal{P}$, with
$\mathcal{P}= \{ n_\sigma,~ \mathcal{R},~
\Theta \}$ 
($\mathcal{R}$ and $\Theta$ refer to the $R_n$ and $\sin^2\theta_W$
nuisance parameters, while $n_\sigma$ stands for the ratio between the experimentally measured CE$\nu$NS cross
section and its SM theoretical value)
\begin{align}
  \label{eq:likelihood}
  \mathcal{L}(m_\chi,\sigma_{\chi-n},\Phi,\mathcal{P})=&
  \prod_{i=1}^{n_\text{bins}} P(N_\text{Exp}^i,N_\text{Obs}^i)
  G(\mathcal{P}_i,\mu_{\mathcal{P}_i},\sigma_{\mathcal{P}_i})
  \nonumber\\
  & \times \prod_{\alpha=1}^{n_\nu}
  G(\phi_\alpha,\mu_\alpha,\sigma_\alpha)\ ,
\end{align}
with $\Phi=(\phi_1, \cdots ,\phi_{n_\nu})$. For the data-driven
analysis $n_\text{bin}$ is dictated by COHERENT data, so
$n_\text{bin}=12$ for CsI and $n_\text{bin}=3$ for LAr
\cite{Akimov:2017ade,COHERENT:2020iec}. For the remaining analyses we consider
$n_\text{bins}=100$. $P(x,n)$ and $G(x,\mu,\sigma)$ are Poisson and
Gaussian probability distribution functions, respectively. This means
that $N_\text{Obs}$ is assumed to be a Poissonian random variable and
that the nuisance parameters follow instead Gaussian distributions
that parametrize their uncertainties. The means $\mu_\alpha$ and
standard deviations $\sigma_\alpha$ are given by the normalization
factors and uncertainties shown in Tab. \ref{tab:parameters}, while
those for the nuisance parameters in the set $\mathcal{P}$ are given
in Tab. \ref{tab:nuisances_mean_variance} and
Fig. \ref{fig:CEvNS_xsec_CsI_LAr}. Note that in the likelihood
function in Eq.(\ref{eq:likelihood}) the Gaussian factors associated
with the nuisance variables in the set $\mathcal{P}$ are
bin-dependent. This is relevant for the data-driven analysis since in
that case the means and standard deviations are energy dependent. In the
case of uncertainties related with $R_n$ and $\sin^2\theta_W$ there is
a single Gaussian bin-independent distribution which factors out. The
resulting likelihood function in those cases thus resembles that used
in Ref. \cite{OHare:2016pjy}.
\begin{table}[h]
  \centering
  \begin{tabular}{|c|c|c|}\hline
    Parameter ($\mathcal{P}$) & Mean ($\mu$)& Unc. (standard deviation)
    \\\hline
    $\mathcal{R}$ & 4.78 fm & 10\%\\\hline
    $\Theta$ & 0.2387 & 10\% \\\hline
  \end{tabular}
  \caption{Mean and standard deviation for the nuisance parameters
    associated with the point-neutron mean-square-radius $R_n$ and
    weak mixing angle $\sin^2\theta_W$ analyses.}
  \label{tab:nuisances_mean_variance}
\end{table}

To set discovery limits one defines a null hypothesis $H_0$ (CE$\nu$NS
background only) and an alternative hypothesis $H_1$ which involves
the WIMP signal plus the CE$\nu$NS background. The likelihood function
in Eq.(\ref{eq:likelihood}) is then specialized for the two cases,
$\mathcal{L}_0$ for $H_0$ and $\mathcal{L}_1$ for $H_1$. In both,
$N_\text{Obs}^i=\sum_\alpha N_\nu^i(\Phi_\alpha, \mathcal{P}_i) + N_W^i(\mathcal{P}_i)$\footnote{$N_W$ involves nuisances only in the case of varying $R_n$.} (see
Eqs.(\ref{eq:N_nu}) and (\ref{eq:N_W})), where $N_\text{Obs}^i$
refers to the total number of ``observed'' events in the $i$-th bin in
a toy experiment defined by a parameter space point
$\{m_\chi, \sigma_{\chi-n}\}$ and neutrino flux normalization factors
as well as $\mathcal{P}$ fixed to their means (see
Tabs. \ref{tab:parameters}, \ref{tab:nuisances_mean_variance} and
Fig.~\ref{fig:CEvNS_xsec_CsI_LAr}). For the generation of these toy
experiments we scan over a $200\times 200$ grid with
$m_\chi\subset [10^{-2},10^3]\,$GeV and
$\sigma_{\chi-n}\subset [10^{-50},10^{-40}]\,\text{cm}^2$.  Note that
the range over which the sum in the first term in $N_\text{Obs}^i$
runs depends on the energy bin. In the first bins all neutrino fluxes
contribute, but as the energy bin increases they start to reach their
kinematic tail and switch off, leaving at $E_r=10^2\,$ keV only the
atmospheric flux contribution.

For $\mathcal{L}_0$ and in the $i$-th bin the ``expected'' number of
events is given by $N_\text{Exp}^i=N_\nu^i(\Phi_\alpha, \mathcal{P}_i)$, with
$N_\nu^i(\Phi_\alpha, \mathcal{P}_i)$ again dictated by Eq.(\ref{eq:N_nu}) but with
the neutrino flux normalizations as well as $\mathcal{P}$ parametrized
in terms of their nuisance variables. For $\mathcal{L}_1$ we make use
of a useful element of the statistical method pointed out by Cowan
\emph{et al.} in Ref. \cite{Cowan:2010js}, namely the Asimov data set. According to the latter,
 it holds that $N_\text{Exp}^i=N_\text{Obs}^i$, while in
the Gaussian factors the nuisance variables are fixed to their
central values. One then calculates $\mathcal{L}_0$ and $\mathcal{L}_1$ for
each parameter space point and then minimizes $-\mathcal{L}_0$ (or
maximizes $\mathcal{L}_0$) for each nuisance variable. For each
parameter space point (toy experiment) one evaluates the likelihood
ratio (test statistics)
\begin{equation}
  \label{eq:likelihood_ratio}
  \lambda(0) = \frac{\mathcal{L}_0}{\mathcal{L}_1}\ .
\end{equation}
This ratio quantifies the disagreement between the null and
alternative hypotheses (or in other words it quantifies the
significance of the WIMP signal), through the \textit{equivalent
  significance} defined according to $Z=\sqrt{-2\,\ln\lambda(0)}$. The
\textit{discovery limit} then follows by finding \textit{the smallest
  WIMP cross section for which $90\%$ of experiments have a WIMP
  signal above $3\sigma$}. In terms of the equivalent significance
this translates into $Z\geq 3$.

With these ingredients we are now in a position to proceed with the
discussion of the effects on discovery limits due to uncertainties on
the measured CE$\nu$NS cross section, $R_n$, $\sin^2\theta_W$ and new
interactions.
\section{WIMP searches}
\label{sec:new-physics}
\subsection{WIMP searches in the presence of standard neutrino
  background}
\label{sec:wimps}
\begin{figure*}
  \centering
  \includegraphics[scale=0.55]{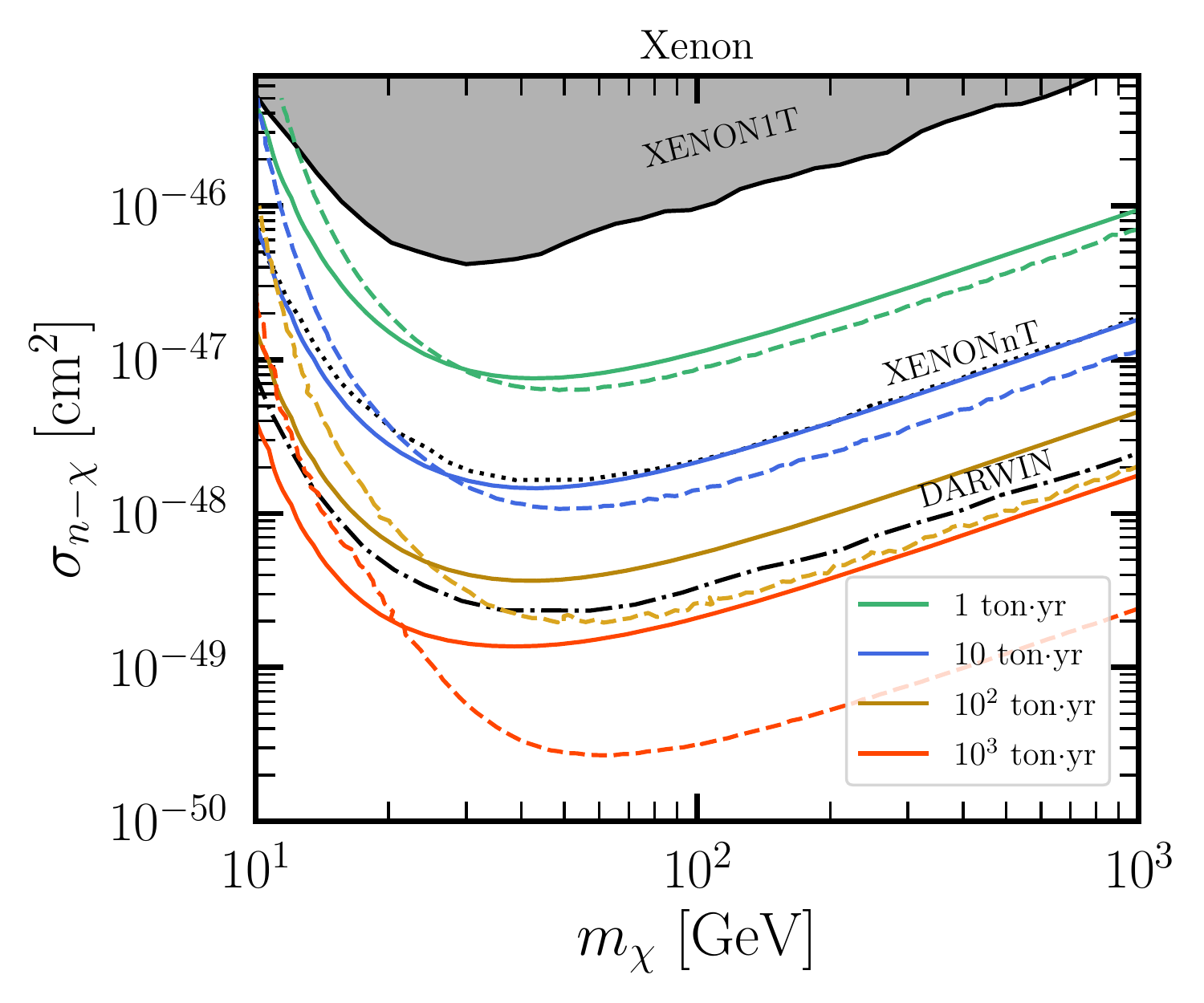}
  \hspace{0.5cm}
  \includegraphics[scale=0.55]{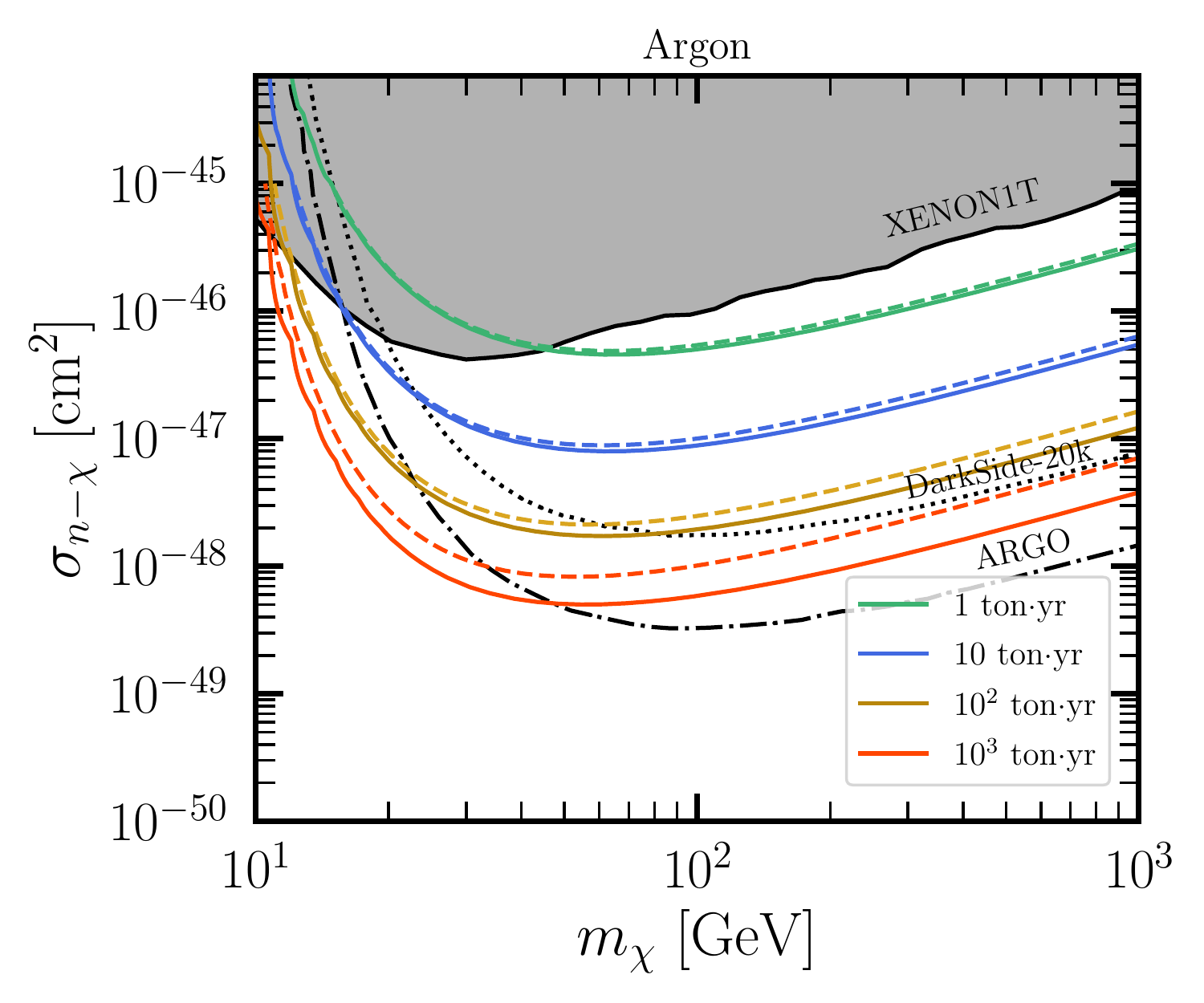}
  \caption{WIMP discovery limits obtained using the CE$\nu$NS cross
    section measurements at COHERENT with the CsI (left graph) and
    LAr (right graph) detectors \cite{Akimov:2017ade,COHERENT:2020iec}
    (dashed curves). In addition to the nuisance parameters due to
    uncertainties on the neutrino flux normalizations, the results
    include bin-dependent nuisance parameters associated with the
    CE$\nu$NS cross section uncertainty as shown in
    Fig.~\ref{fig:CEvNS_xsec_CsI_LAr}. The current constraint set by XENON1T is shown in both panels. Moreover, we show as for comparison future sensitivities expected at LXe experiments XENONnT and DARWIN (left panel)  \cite{XENON:2020kmp,Aalbers:2016jon} and at LAr experiments DarkSide-20k and ARGO (right panel) ~\cite{DarkSide-20k:2017zyg,Billard:2021uyg}.}
  \label{fig:data-driven}
\end{figure*}
\subsubsection{Data-driven analysis}
\label{sec:data-driven}
In the ``standard'' calculation of WIMP discovery limits the CE$\nu$NS
cross section is assumed to be known with $100\%$ accuracy
\cite{Billard:2013qya}. Uncertainties due to nuclear form factors or
other quantities such as the weak mixing angle are not
considered. Where the uncertainty on the cross section resides or
whether there is new physics contributing to it---the subject of
Sec. \ref{sec:wimp-and-new-bckg}---is to a certain extent an
assumption-dependent question. To avoid this a
data-driven approach can be rather adopted, in which one uses the measured CE$\nu$NS
cross section along with its uncertainty. This approach encapsulates
all possible uncertainties the cross section can involve, regardless
of assumptions.

To proceed, we first extract from the COHERENT CsI and LAr data the
CE$\nu$NS cross section central values along with their standard
deviations (the CsI data are directly applicable to xenon since both
nuclides have about the same average mass and atomic numbers). To do
so, we weigh the theoretical SM value of the CE$\nu$NS differential cross section with a
multiplicative factor $n_\sigma$ and use a spectral $\chi^2$ test to
fit $n_\sigma$ in each recoil energy bin (see App. \ref{sec:extraction} for
details). Assuming that the CE$\nu$NS differential cross section
uncertainty is fully encoded in a multiplicative factor is the most simple
approach one can adopt. Given the quality of the data sets, the
uncertainty could be assumed to be energy dependent. However, modeling
such an energy-dependent uncertainty seems to us more arbitrary (there is
a few number of functions one could use) than assuming a flat
uncertainty.

For the data-driven analysis with COHERENT CsI data we use 12 bins starting at 7 photoelectrons (PE)
and extending up to 29 PE ($\text{PE}=1.17\,E_r/\text{keV}_\text{nr}$)
\cite{Akimov:2017ade}, while for the LAr dataset we use three bins starting
at  $5\,\text{keV}_\text{ee}$ and up to $25\,\text{keV}_\text{ee}$
($\text{keV}_\text{nr}\simeq \text{keV}_\text{ee}/4$)
\cite{COHERENT:2020iec}. Indeed, for energies above $25\,\text{keV}_\text{ee}$
the CE$\nu$NS event rate is small enough and the remaining bins are of
little relevance. Note that in the definition of the $\chi^2$ test, to extract the $n_\sigma$ factors, 
systematic errors associated to neutrino flux and form factor uncertainties have been included as
nuisance parameters. The results presented in
Fig. \ref{fig:CEvNS_xsec_CsI_LAr} thus encode only uncertainties due
to the cross section (indirect) measurement.

Exploiting this determination of the uncertainties 
on the CE$\nu$NS cross section from COHERENT data we then compute the WIMP
discovery limits. We use the general
definition of the likelihood function in Eq.(\ref{eq:likelihood}) along
with the results depicted in Fig. \ref{fig:CEvNS_xsec_CsI_LAr}. This implies
that the regions that can be covered correspond to those affected by
DSNB and sub-GeV atmospheric neutrino backgrounds (heavy WIMP
masses). The results are displayed in Fig. \ref{fig:data-driven}, using CsI (LAr) data in the left (right) panel.  In the analysis with CsI data one can see that in general, compared with the SM
expectation (solid curves), WIMP discovery limits improve. A closer
inspection to the left graph in Fig. \ref{fig:CEvNS_xsec_CsI_LAr}
allows to understand this behavior. Except for bin number 12, the measured CE$\nu$NS
cross section (central values) is smaller than the SM expectation,
thus resulting in a background depletion which becomes more visible
with increasing exposure. However, for low WIMP masses, $m_\chi \lesssim 20$~GeV, our likelihood analysis tends to favor the maximum cross section values, hence leading to a poorer sensitivity compared to the pure SM case. 

Results derived using the LAr data behave instead the other way
around. The data trend is that of a measured CE$\nu$NS cross section exceeding its SM
expectation, as can be seen in the right graph of
Fig. \ref{fig:CEvNS_xsec_CsI_LAr}. Departures, however, are not
substantial and thus the enhancement of the background is not that
large. As a result, discovery limits are only slightly worsen as
shown in the left graph in Fig. \ref{fig:data-driven}.

Further data from CE$\nu$NS experiments will allow an improvement
of the discovery limit we have presented here. Larger statistics
combined with a better understanding of systematic and statistical
uncertainties (as expected in future reactor and COHERENT
experiments), will allow for a less spread CE$\nu$NS cross section
measurement. However, even if statistics becomes abundant and
uncertainties are shrinked to values below the percent level, neutrino
flux uncertainties will still affect the exact position and shape of
the neutrino floor. Improvements of the discovery limit will require
therefore not only better measurements of the CE$\nu$NS cross section
but of relevant neutrino fluxes, in particular those from the $^8$B
component.

\begin{figure*}
  \centering
  \includegraphics[scale=0.55]{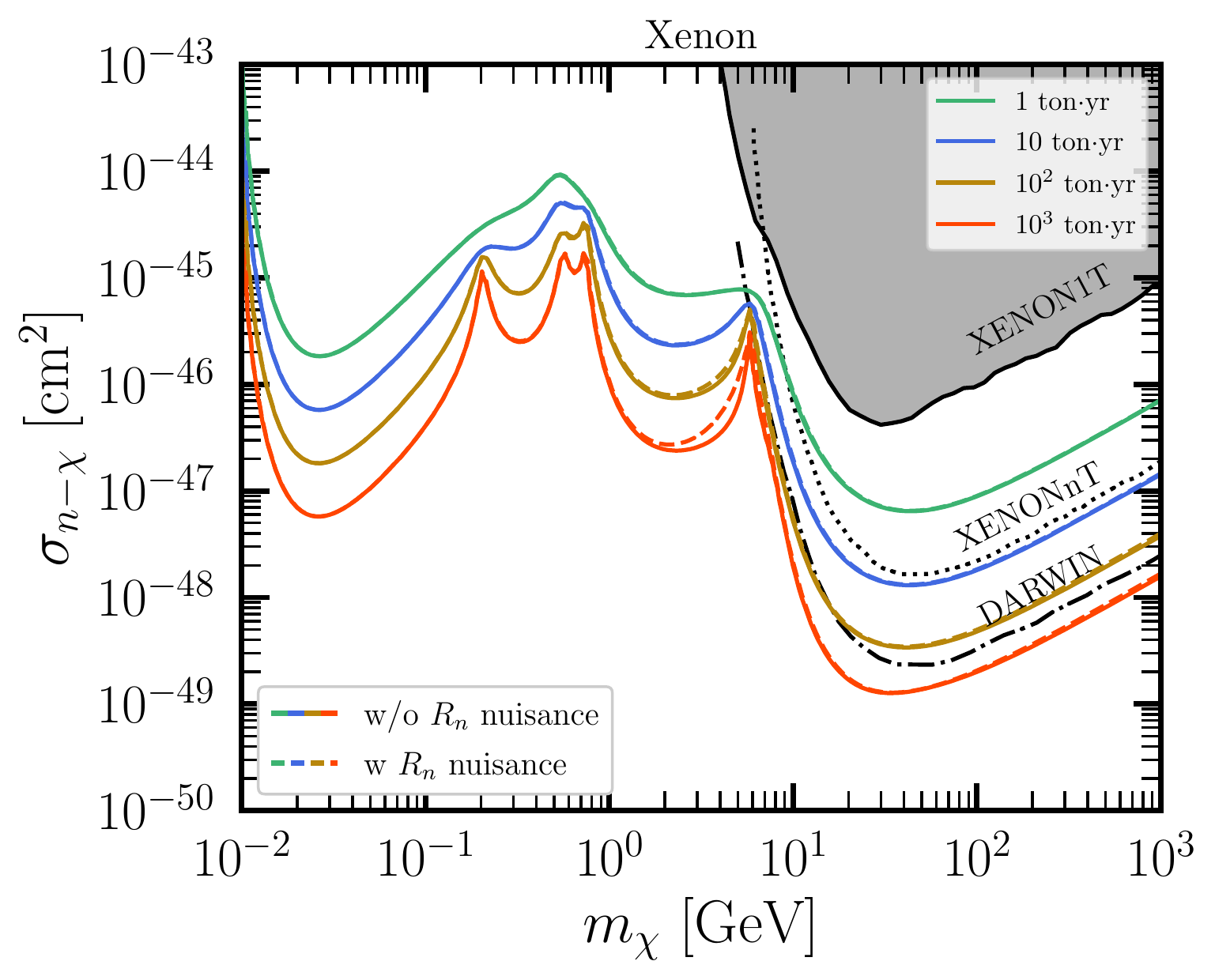}
  \hspace{0.5cm}
  \includegraphics[scale=0.55]{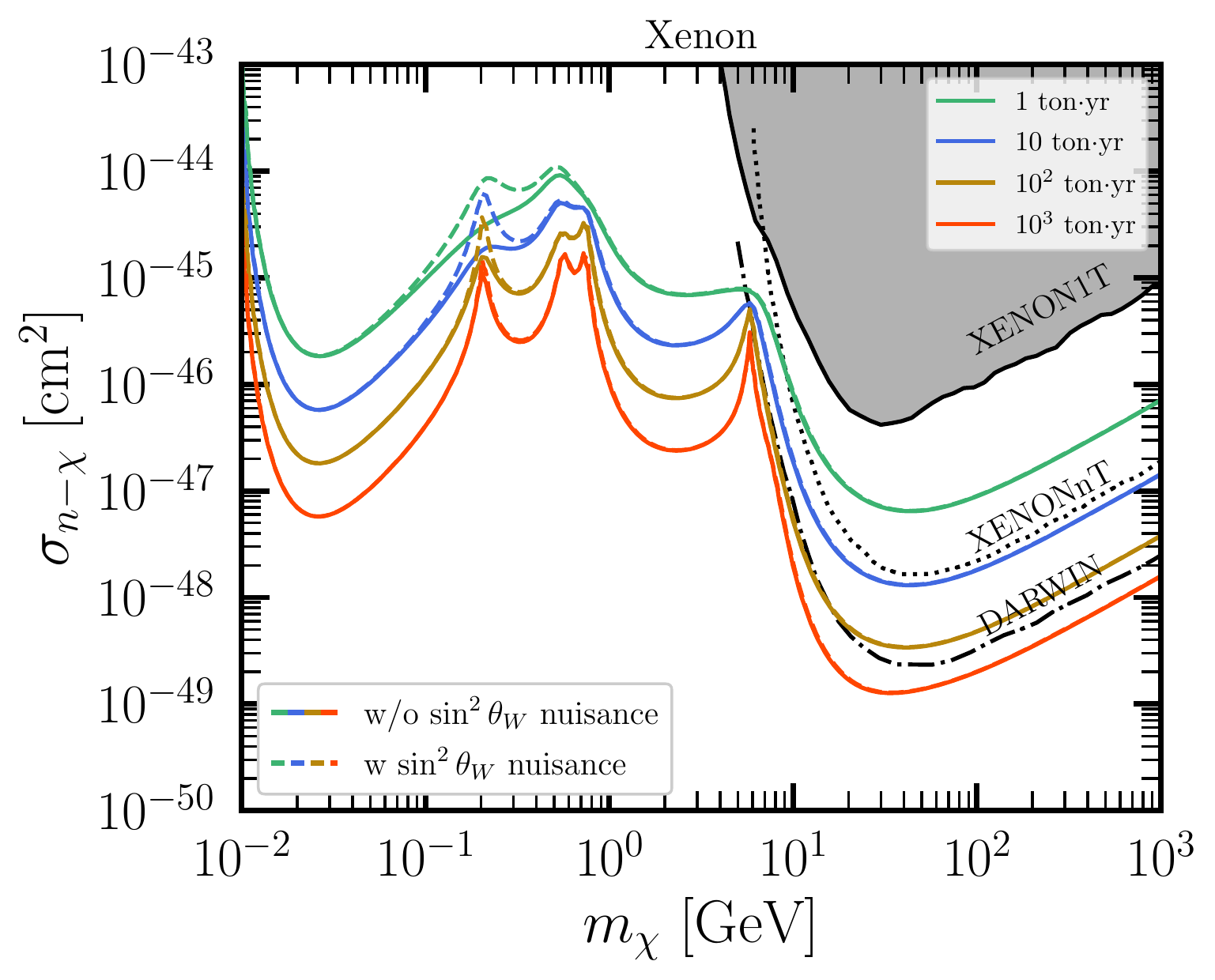}
  \caption{\textbf{Left graph}: WIMP discovery limit calculated by
    considering neutrino flux uncertainties as well as uncertainties
    on the xenon neutron distribution mean-square-radius. Results are
    presented for four different exposures and are compared with
    results obtained solely by considering neutrino flux uncertainties
    (solid curves). XENON1T, XENONnT and DARWIN sensitivities are
    shown for comparison
    \cite{Aprile:2015uzo,Aalbers:2016jon,XENON:2020kmp}. \textbf{Right graph}: WIMP
    discovery limits calculated by considering uncertainties on the
    weak mixing angle at low energies. }
  \label{fig:neutron_weak_mix_angle}
\end{figure*}
\subsubsection{Impact of nuclear form factor and weak mixing angle
  uncertainties}
\label{sec:impact-quenching-FF}
In contrast to proton distributions, neutron distributions are poorly
known. With the exception of neutron distributions for
$^{208}$Pb---measured with high accuracy using parity violating
electron scattering by the PREX experiment
\cite{Abrahamyan:2012gp,HAPPEX:2012fud}---and more recently for
$^{133}$Cs and $^{127}$I using COHERENT data
\cite{Cadeddu:2017etk,Cadeddu:2021ijh}, little is known about neutron
distributions for other nuclei. Given the typical incoming neutrino
energies for which CE$\nu$NS can be observed, nuclear effects are not
as sizable as they are for other neutrino scattering processes such as
e.g. quasi-elastic scattering or resonant pion production (see
e.g. \cite{Formaggio:2013kya}). This, however, does not mean that they
can be fully ignored.

Differences between proton and neutron distributions are expected to
be substantial for neutron-rich nuclei. These departures have in turn
an impact on the values the nuclear form factor can have at a
particular momentum transfer \cite{AristizabalSierra:2019zmy}. At very
low $q$---typical of reactor or solar neutrino fluxes---deviations are
small and so have little impact on the neutrino event spectrum. As $q$
increases to larger values---typical of stopped-pion or sub-GeV
atmospheric neutrino fluxes---uncertainties on the point-neutron
distribution mean-square radius become relevant and can have an
impact. Note that this effect applies as well to the WIMP event rate,
so not only the CE$\nu$NS event spectrum comes along with an
uncertainty but also the WIMP event spectrum. However, in contrast to
the WIMP rate, which involves as well uncertainties due to
astrophysical parameter inputs, nuclear physics uncertainties entirely
determine the precision with which CE$\nu$NS can be precisely
predicted.

The weak-charge form factor in Eq.(\ref{eq:CEvNS-xsec}) can be
written according to \cite{Coloma:2020nhf}
\begin{align}
  \label{eq:EW-FF-approx}
  F_W&\simeq \frac{1}{Q_\text{W}}
       \left[
       Z\left(
       g_V^p 
       - \frac{g_V^p}{6}r_p^2 q^2 
       - \frac{g_V^n}{6}r_n^2 q^2
       \right) F_p(q^2)\right .
       \nonumber\\
     &\left .
       +
       N\left(
       g_V^n 
       - \frac{g_V^n}{6}r_p^2 q^2 
       - \frac{g_V^p}{6}r_n^2 q^2
       \right) F_n(q^2)
       \right]\ ,
\end{align}
where the spin-independent proton and neutron nuclear form factors are
normalized, $F_p(q^2=0)=1$ and $F_n(q^2=0)=1$. The quantities in front of
the nuclear form factors are the leading-order nucleon form factor
terms. Given the momentum transfer values we are dealing with, in our
analysis we keep only the $q^2$-independent terms. Including order
$q^2$ terms will correct our results at the percent level in the
atmospheric neutrino region (heavy WIMP mass region), which, given the
expected neutrino event rate (statistics), is too small to yield a
sizable effect \cite{daristi:2021}. Note that in
Eq.(\ref{eq:WIMP-N_WIMP-nucleon_xsec}) the same assumption has been
adopted, in addition to the assumption of a WIMP isospin-conserving
interaction.

As we have pointed out, we adopt the Helm parametrization for the spin-independent proton and
neutron nuclear form factors. It relies on
the assumption that nucleon distributions follow from a convolution of
an uniform distribution of radius $R_0$ (diffraction radius) and a
Gaussian profile, characterized by the folding width $s$ that accounts
for the surface thickness. The Helm form factor is then given by
\begin{equation}
  \label{eq:helm-FF}
  F(q^2)=3\frac{j_1(qR_0)}{qR_0}\,e^{-\frac{1}{2}(qs)^2}\ ,
\end{equation}
where $j_1(x)$ is a spherical Bessel function of the first type and
$s=0.9\,$fm \cite{Collar:2019ihs}. The diffraction radius is
determined by the mean-square radius of the corresponding distribution
according to
\begin{equation}
  \label{eq:diffraction_radius_Helm_parametrization}
  R_0 = \sqrt{\frac{5}{3}\left(R_X^2-3s^2\right)}\qquad (X=p, n)\ .
\end{equation}
Compared with the form factor parametrization based on the symmetrized
Fermi distribution and the Klein-Nystrand parametrization
\cite{Piekarewicz:2016vbn,Klein:1999qj}, the Helm parametrization
tends to underestimate event rates
\cite{AristizabalSierra:2021uob}. For momentum transfer values as
those implied by sub-GeV atmospheric neutrino fluxes (heavy WIMP
masses), event rates interpolate between values determined by the Helm
and the Klein-Nystrand form factors with variations of order
$10\%$. So, for definiteness we choose to work with the Helm
parametrization understanding that event rates for both WIMP and
neutrinos fluctuate by about $10\%$.

In the statistical analysis, we calculate $N_\text{Obs}^i$ by fixing
$R_p$ to its averaged value calculated according to
$R_p=\sum_{i=1}^9 X_i R_p^i$, where $X_i$ refers to the relative
abundance of the $i$-th xenon stable isotope and $R_p^i$ to its proton
distribution root-mean-square radius extracted from
Ref. \cite{Angeli:2013epw}.  We obtain the averaged value
reported in Tab.~\ref{tab:nuisances_mean_variance}. For the averaged
neutron distribution root-mean-square radius we use
$R_n=R_p=\mathcal{R}$. We then allow $10\%$ variations above this
value in the Poisson and Gaussian factors in $\mathcal{L}_0$. We then
calculate the equivalent significance for each parameter space point
and after imposing $Z\geq 3$ we get the discovery limit determined not
only by neutrino flux uncertainties but also by uncertainties on the
xenon neutron distribution mean-square radius (nuclear form factor
uncertainty).

The result is shown in the left graph of
Fig.~\ref{fig:neutron_weak_mix_angle}. One can see that for low WIMP
masses there is no difference between the result obtained by
considering neutrino flux uncertainties alone (solid curves) and the result
including additional neutron distribution mean-square radius uncertainties
(dashed curves). This is expected since for those WIMP masses and
incoming neutrino energies, the zero momentum transfer limit is a
rather good approximation. Form factor effects are thus negligible. As
the WIMP mass increases and so the incoming neutrino energy, the
effect starts to show up and becomes more pronounced as
exposure increases. The region in WIMP mass where sizable deviations
are observed corresponds to regions where the dominant background is
determined by $^8$B and hep neutrino fluxes. In that region the
typical momentum transfer is of the order of 30-40 MeV, for which the
form factor spread is of the order of 2-3\%
\cite{AristizabalSierra:2019zmy}. At low exposure, the size of the
uncertainty combined with a mild statistics leads to an effect which
can be barely noted. However, as exposure increases (and so
statistics) the effect becomes more prominent.

The variation of the WIMP discovery limit due to the form factor
uncertainty is also expected. With increasing neutron mean-square radius, nuclear size
increases. A larger nuclear radius in turn means that the loss of coherence
happens for smaller $q$. As $R_n$ increases, up to the $10\%$
above $R_p$, the number of neutrino events decreases and so the WIMP
event rate. The overall effect is that of a diminished discovery limit
at high exposures, though rather feeble. For increasing WIMP masses
and neutrino energies (entering the atmospheric neutrino region), form
factor uncertainties increase up to the order of $5\%$, but statistics
becomes scarce even for the maximum exposure that we have chosen. As a
result, the discovery limit at high WIMP masses becomes insensitive to
this effect. One can fairly conclude that uncertainties on the WIMP
and neutrino event rates originating from the xenon neutron
distribution mean-square radius have little impact on the WIMP
discovery limit. The reason behind this behavior is first of all
related with the fact that nuclear form factor uncertainties in xenon
are \textit{per se} small \cite{AristizabalSierra:2019zmy}. It is
secondly related with another fact: once $R_n$ increases, both the neutrino
background and the WIMP event rate are (slightly) suppressed.

We now turn to the discussion of uncertainties on the low-energy value
of the weak mixing angle. The best measurement of $\sin^2\theta_W$ has
been done at the $Z$ scale at Tevatron, LEP and more recently at the
LHC (see e.g.~\cite{ParticleDataGroup:2018ovx}). The precision of
those measurements is of the order of $\sim 0.1\%$. At other (lower)
scales---that span about five orders of magnitude---measurements
include NuTeV, parity violation in electron scattering, electron and
proton weak charge and atomic parity violation
\cite{Zeller:2001hh,Wang:2014bba,Anthony:2005pm,Androic:2018kni,Wood:1997zq}. In
contrast to measurements at the $Z$ scale, these measurements involve
order $\sim 10\%$ uncertainties.

Lacking the precision of accelerator experiments, these measurements
still allow variations of the weak mixing angle that can have an
impact on WIMP discovery limits. However, for this to be the case a
large CE$\nu$NS statistics is required given that $g_V^p\ll
g_V^n$.
This means that effects of weak mixing angle uncertainties are
expected to be relevant at low WIMP masses, where solar neutrino
fluxes are more abundant.

As in the nuclear form factor case, this effect is also---in
principle---energy dependent. At each WIMP mass there is a neutrino
flux that matters, and therefore a $q_\text{max}=\sqrt{2m_NE_r^\text{max}}$
that fixes the renormalization scale at which $\sin^2\theta_W$ should
be evaluated. In other words, as $m_\chi$ increases the mean value for
$\sin^2\theta_W$ changes since the renormalization scale does
so. There is however an interesting observation that allows for the following simplification: for
the renormalization scales ($q_\text{max}$) that matter in the
calculation of WIMP discovery limits
($q_\text{max}\lesssim 200\,$MeV), the weak mixing angle RGE evolution
(in the SM) is rather flat \cite{Erler:2004in}. One can then fix its
mean value to its zero momentum transfer value obtained by
extrapolation \cite{Kumar:2013yoa}
\begin{equation}
  \label{eq:weak_mix_angle_extrapolation}
  \sin^2\theta_W(q=0)=\kappa(q=0)_{\overline{\text{MS}}}
  \sin^2\theta_W(m_Z)_{\overline{\text{MS}}}\ ,
\end{equation}
where the parameter at $q=0$ is given by
$\kappa(q=0)_{\overline{\text{MS}}}=1.03232\pm 0.00029$ and the weak
mixing angle at the $Z$ scale by
$\sin^2\theta_W(m_Z)_{\overline{\text{MS}}}=0.23122\pm 0.00003$
\cite{ParticleDataGroup:2018ovx}. Taking the central values and
allowing for a $10\%$ uncertainty one can then calculate the WIMP
discovery limits obtained by combining neutrino flux normalization and
weak mixing angle uncertainties. The result is displayed in the right
graph of Fig. \ref{fig:neutron_weak_mix_angle}. As expected, the
effect of the weak mixing angle uncertainty becomes visible at low
WIMP masses. The region where sizable deviations from the ``standard''
result are more pronounced corresponds to those where the pp, $^7$Be
(two lines) and $^{13}$N dominate the background. Note that once less
abundant neutrino fluxes kick in (from $^8$B on) the discovery limit
converges to the ``standard'' result. The reason is the combination of a small effect and low statistics.

Overall, the behavior of the discovery limit in the presence of this
uncertainty can be readily understood. As the weak mixing angle
increases, the coherent weak charge becomes more
negative. Quadratically, the coupling increases (decreases) about $\sim 12\%$ when
allowing the weak mixing angle to increase (decrease) by $10\%$. Although the
enhancement is not dramatically large, it is sufficient to increase
the number of neutrino events. 
\begin{figure*}
  \centering
  \includegraphics[scale=0.55]{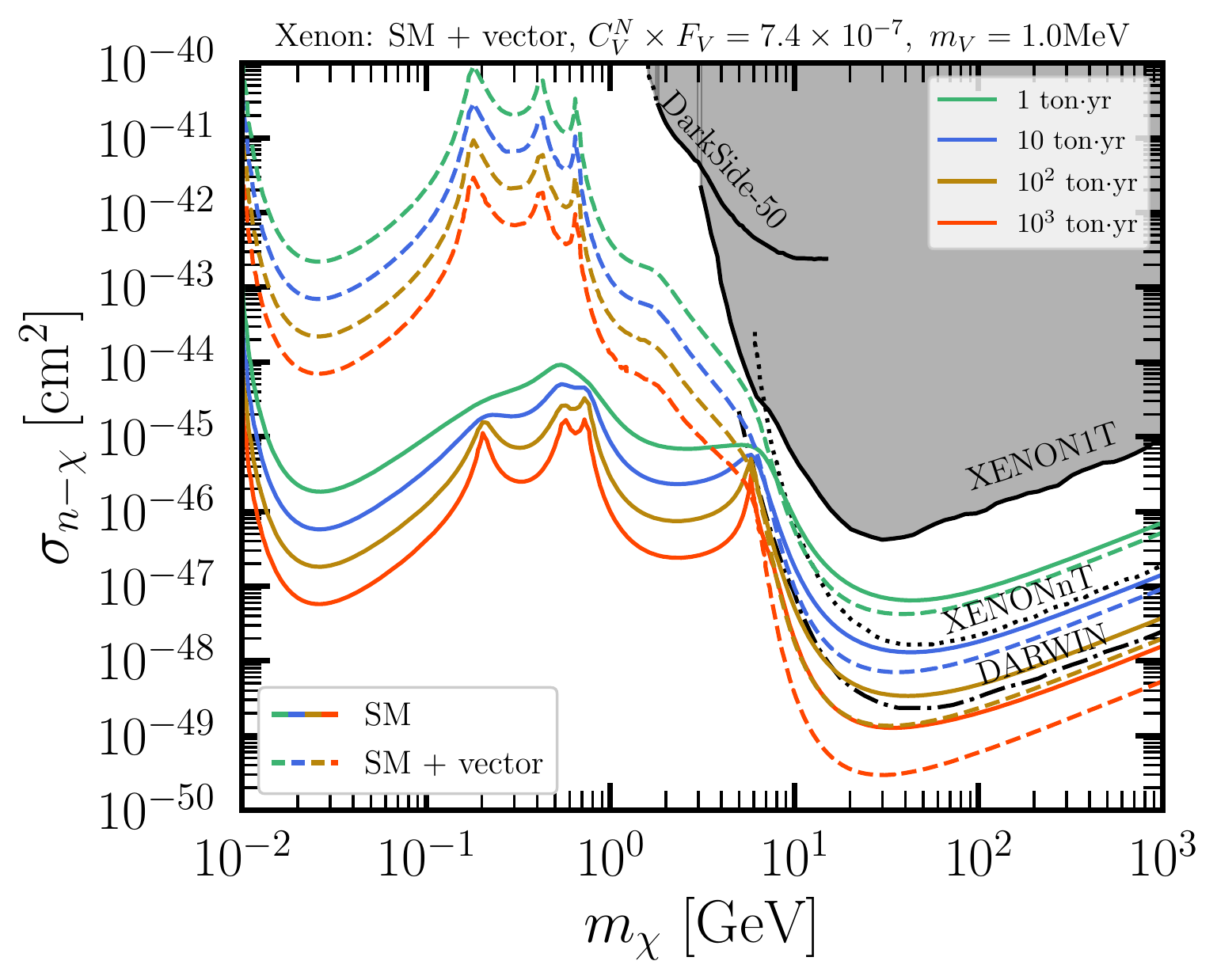}
  \hspace{0.5cm}
  \includegraphics[scale=0.55]{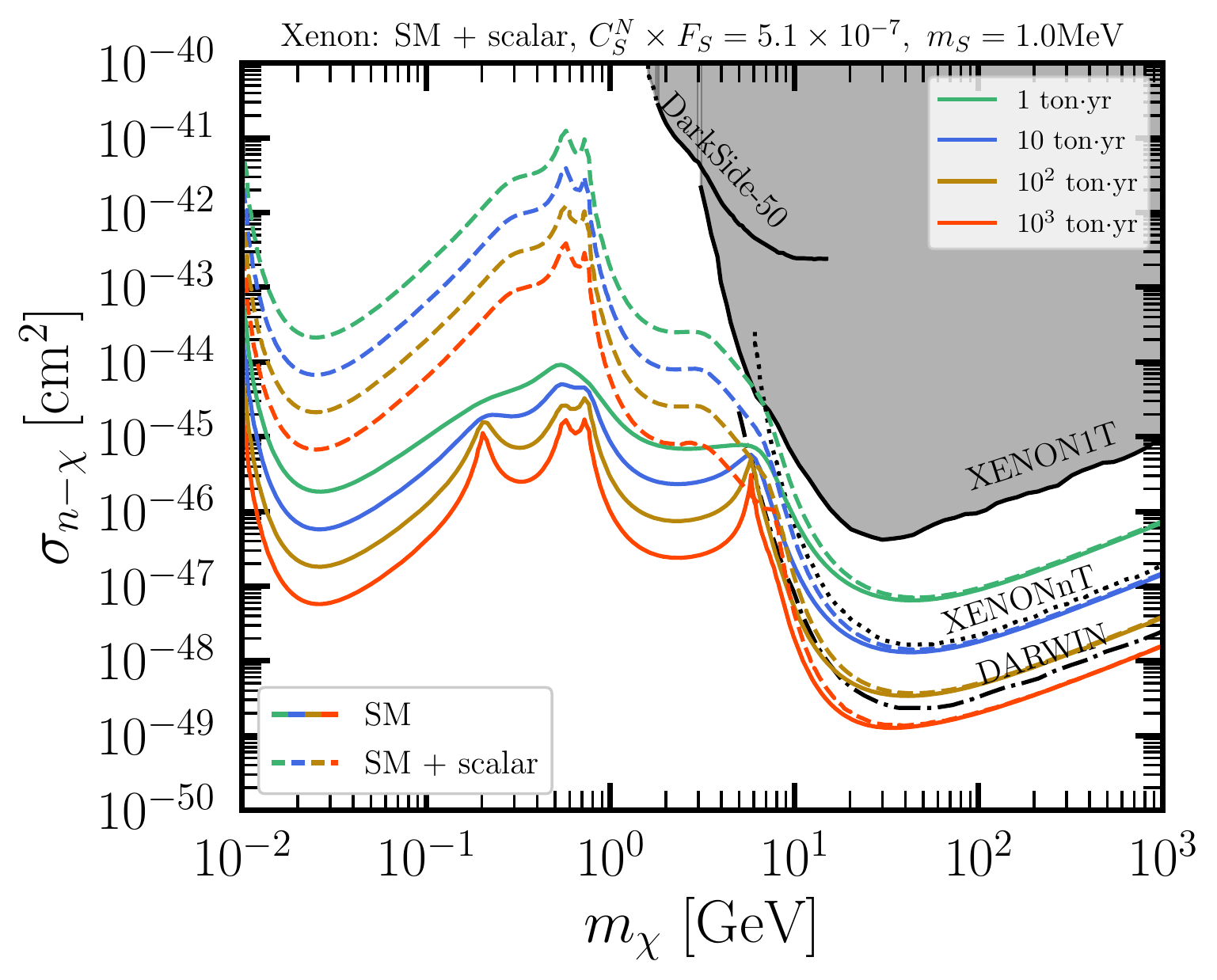}
  \caption{\textbf{Left graph}: WIMP discovery limit in the presence
    of a long-range vector interaction calculated for four different
    exposures and for values of the coupling and vector boson mass fixed to maximize
    its effect. Along with the result (dashed curves), the SM
    discovery limits (solid curves) are shown for
    comparison. \textbf{Right graph}: Same as for left graph but for a
    long-range scalar interaction. Couplings and masses have been
    fixed as required by COHERENT CsI data
    \cite{AristizabalSierra:2019ykk}, they correspond to the 90\% CL
    upper bounds. In both panels we also show for comparison the upper limits set by XENON1T and DarkSide-50 \cite{Aprile:2015uzo,DarkSide:2018kuk,DarkSide:2018bpj}, together with future sensitivities for XENONnT and DARWIN  \cite{Aalbers:2016jon,XENON:2020kmp}.}
  \label{fig:vector_scalar}
\end{figure*}
\subsection{WIMP searches in the presence of non-standard neutrino
  background}
\label{sec:wimp-and-new-bckg}
As far as we know, new physics in neutrino backgrounds at multi-ton DM
detectors have been discussed using the so-called \textit{one neutrino
  event contour line} in
Refs. \cite{Dutta:2017nht,Bertuzzo:2017tuf,AristizabalSierra:2017joc,Boehm:2018sux,Essig:2018tss,AristizabalSierra:2019ykk}. Analyses
aiming at determining the impact of the new interactions on WIMP
discovery limits have been instead presented in
Ref. \cite{Bertuzzo:2017tuf,Gonzalez-Garcia:2018dep,Wyenberg:2018eyv,Gaspert:2021gyj}, using neutrino nonstandard
interactions (NSI) as the new physics contribution. Here we first revisit
results for interactions including light vector and scalar mediators,
and then we present new results considering neutrino magnetic moments/transitions. 

In the light of COHERENT data and of other forthcoming CE$\nu$NS
experiments
\cite{Akimov:2017ade,COHERENT:2020iec,Bonet:2020awv,Aguilar-Arevalo:2019jlr,Strauss:2017cuu,SBC:2021yal},
light vector mediators have been the subject of recent analyses (see
e.g. \cite{Liao:2017uzy,AristizabalSierra:2019ufd}). In the presence
of a new vector lepton flavor-conserving interaction the CE$\nu$NS
differential cross follows from Eq.(\ref{eq:CEvNS-xsec}) by shifting
the coherent weak charge $Q_W$ according to
\begin{equation}
  \label{eq:weak_charge_shifted_vector}
  Q_V = Q_W + \frac{C_V^NF_V}{\sqrt{2}G_F(2m_NE_r+m_V^2)}\ ,
\end{equation}
where $m_V$ is the mass of the vector mediator while the coupling $F_V$
determines the strength at which the vector couples to neutrinos
through vector and axial couplings. The coupling $C_V^N$, instead,
determines the coupling of the new vector boson to the nucleus
\begin{equation}
  \label{eq:CVN}
  C_V^N = (A-Z)(h_V^u+2h_V^d) + Z(2h_V^u+h_V^d)\ ,
\end{equation}
with $h_V^q$ the vector current couplings of the vector boson to up
and down quarks (compared with the vector current, the axial current
is suppressed and so no axial couplings are included). From the
structure of the new weak charge, it is clear that the new interaction
can either constructively or destructively interfere with the SM
contribution. The presence of a new vector interaction can therefore
enhance or deplete the SM contribution and depending on the momentum
transfer, $q^2=2m_NE_r$, can lead to spectral distortions as well
\cite{AristizabalSierra:2019ufd,Abdullah:2020iiv}.

In order to maximize the effects implied by the new interaction we fix
the product of couplings $C_V^NF_V$ to their maximum allowed value
according to COHERENT CsI data at $m_V=1\,$MeV
\cite{AristizabalSierra:2019ykk}: $C_V^NF_V\lesssim 7.4\times 10^{-7}$
at the $90\%\,$CL \footnote{Statistically the observation of CE$\nu$NS
  with the CsI detector is more robust than with the LAr detector. In
  the former, data favors the observation of the signal over
  background at the $6.7\sigma\,$CL, while in the latter at the
  $3\sigma\,$CL \cite{Akimov:2017ade,COHERENT:2020iec}. That is why we
  use constraints derived using CsI data
  \cite{AristizabalSierra:2019ykk}.}. Note that this value applies to
CsI, but can be used for xenon as well given the similarity of these
nuclides. Light vector mediators are subject to constraints that
follow from stellar cooling arguments and neutrino diffusion time
disruption in supernova environments
\cite{Grifols:1986fc,Grifols:1988fv,Chang:2016ntp}. The combination of
coupling and mass that we have chosen is reconcilable with these bounds.

We then
calculate the WIMP discovery limit. In this case the only nuisance
parameters are those associated with neutrino flux normalization
factors. However, depending on the transfer momentum, the new
contribution can enhance (deplete) the neutrino background thus
worsening (improving) the discovery limit. This is confirmed by the result shown in the
left graph of Fig. \ref{fig:vector_scalar}.

\begin{figure*}
  \centering
  \includegraphics[scale=0.6]{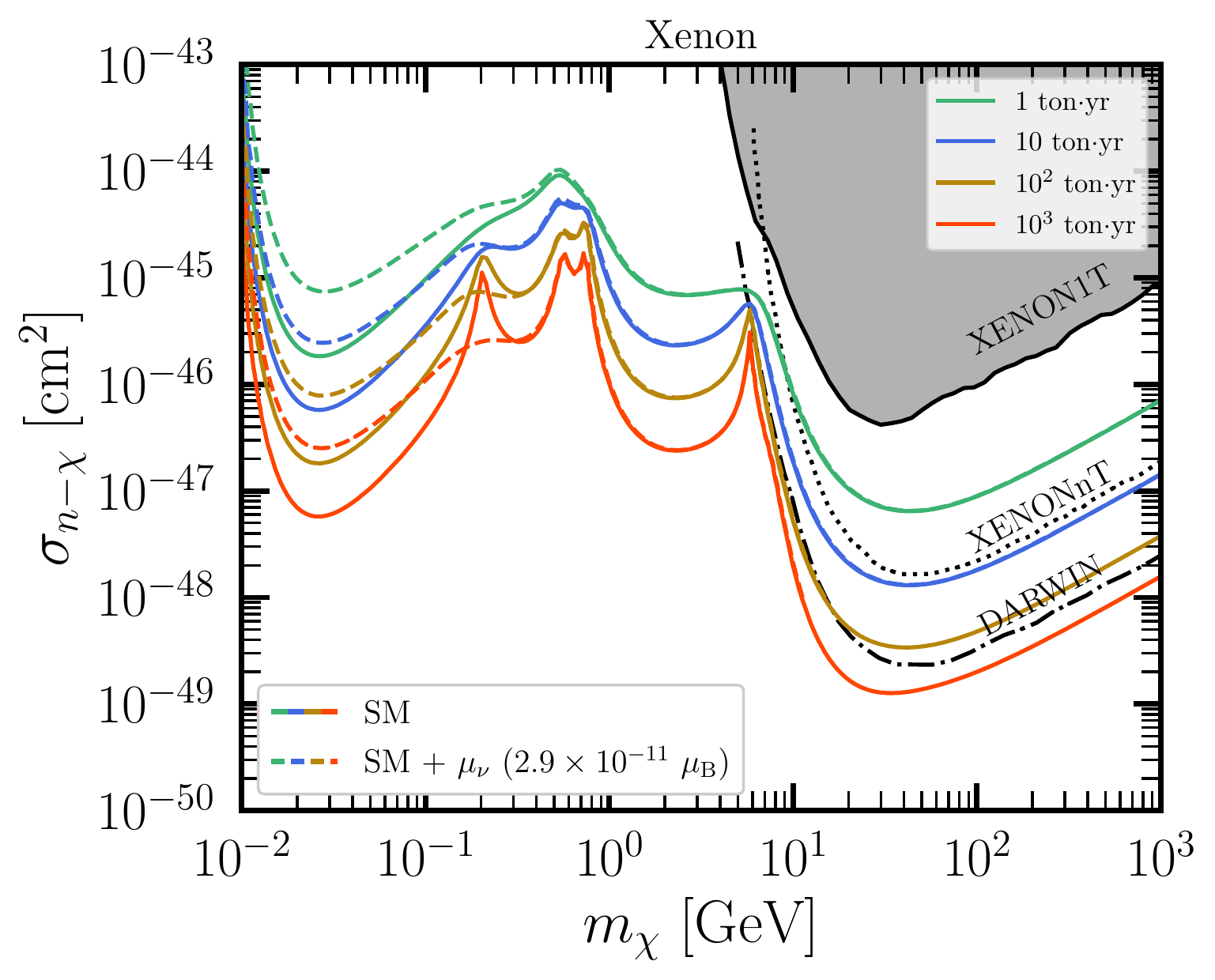}
  \caption{WIMP discovery limits in the presence of neutrino
    magnetic/transition interactions along with discovery limits in
    the SM alone. The neutrino magnetic moment has been fixed to
    $2.9\times 10^{-11}\mu_B$, the $90\%\,$CL upper limit reported by
    the GEMMA reactor experiment \cite{Beda:2010hk}.}
  \label{fig:nmm}
\end{figure*}
At low WIMP masses the discovery limit is diminished by several orders
of magnitude. This can be readily understood by the $q^2$ dependence
of the new contribution. At low momentum transfer this term is
enhanced and overcomes the SM contribution, as the CE$\nu$NS cross
section is enhanced towards low momentum transfer regions. As a result,
the neutrino background increases, thus leading to a dramatic
diminishing of the discovery limit. For regions above $10\,$GeV, after
the $^8$B and hep neutrino fluxes reach their kinematic tail, the
discovery limit improves. This crossover can be understood as
follows. The SM coherent weak charge is negative, while the new
contribution is positive. So, as $q^2$ increases the new contribution
becomes less prominent and destructively interferes with the SM term,
leading to a suppression of $Q_V$. The background then becomes less
severe, thus resulting in an improvement of the discovery limit.

Scalar interactions in both the effective and light regimes have been
as well recently considered in the context of CE$\nu$NS related
experiments
\cite{Farzan:2018gtr,AristizabalSierra:2018eqm,Brdar:2018qqj}. Since
the scalar coupling involves a chirality flip it cannot (sizably)
interfere with the SM contribution, in contrast to the vector
interaction. Thus, in the presence of the scalar coupling, the CE$\nu$NS
cross section consists of two terms, the SM term in
Eq.(\ref{eq:CEvNS-xsec}) and a second term given by (assuming universal lepton
flavor couplings)
\cite{Cerdeno:2016sfi,Farzan:2018gtr,AristizabalSierra:2019ykk}
\begin{equation}
  \label{eq:scalar_CEvNS}
  \frac{d\sigma_S}{dE_r}=\frac{G_F^2}{2\pi}m_NQ_S^2
  \frac{m_NE_r}{2E_\nu^2}F^2(q^2)\ ,
\end{equation}
where the scalar charge $Q_S$ reads
\begin{equation}
  \label{eq:scalar_charge}
  Q_S=\frac{C_S^NF_S}{G_F(2m_NE_r + m_S^2)}\ .
\end{equation}
Here, $F_S$ measures the strength at which the scalar couples to
neutrinos through scalar and pseudoscalar couplings, and $C_S^N$
determines the coupling of the scalar to the nucleus according to
\begin{equation}
  \label{eq:CSN}
  C_S^N=(A-Z)\sum_{q=u,d}h_S^q\frac{m_n}{m_q}f_{T_q}^n
  + Z\sum_{q=u,d}h_S^q\frac{m_p}{m_q}f_{T_q}^p\ ,
\end{equation}
with $h_S^q$ being the couplings of the scalar to up and down quarks. The
hadronic form factors $f^{n,p}_{T_q}$ follow from chiral perturbation
theory calculations using measurements of the $\pi$-nucleon sigma
term. Their values can be found in
e.g.~\cite{Hoferichter:2015dsa,Ellis:2018dmb}.

Results for the impact of this interaction on WIMP discovery limits
are shown in the right graph of Fig.~\ref{fig:vector_scalar}. As in
the vector case, the product $C_S^NF_S$ has been fixed to its maximum
allowed value at $m_S=1\,$MeV according to COHERENT CsI data:
$C_S^NF_S\lesssim 5.1\times 10^{-7}$ at the $90\%\,$CL. This value is
consistent with astrophysical and cosmological bounds as well as with
bounds derived from neutrino masses, which are generated by the scalar
coupling below $\Lambda_\text{QCD}$
\cite{AristizabalSierra:2019ykk}. The results follow expectations. The
scalar contribution peaks towards the low momentum transfer region
(low WIMP mass region), thus enhancing the background and so worsening
the discovery limit. One can see that the degree at which the
discovery limit is affected is less severe than in the vector
case. This is also expected, since the leading vector term is linear in the
coupling while the scalar contributes quadratically. At high momentum
transfer (high WIMP mass) the scalar keeps enhancing the background, hence
there is no crossover as in the vector case. Since destructive
interference is not possible in the scalar case, the discovery limit is still diminished at $m_\chi \gtrsim 10$ GeV, though
less than in the low WIMP mass region due to the larger momentum
transfer involved.\\

We finally move to the case of neutrino magnetic moments/transitions,
which have been a subject of recent interest in the context of
COHERENT data and multi-ton DM experiments
\cite{Papoulias:2019txv,Miranda:2019wdy,Papoulias:2019xaw,Kosmas:2017tsq,AristizabalSierra:2020zod}. The
neutrino electromagnetic current can be parametrized in terms of four
form factors which in the zero momentum transfer limit define the
neutrino: electric charge, electric dipole moments (or transitions),
magnetic dipole moments (or transitions) and anapole
moments. These parameters enable the coupling of neutrinos to photons
and so through $t$-channel processes they contribute to CE$\nu$NS and
neutrino-electron elastic scattering (see \cite{Giunti:2014ixa} for a
review). The new processes do not interfere with the SM contribution,
so the total CE$\nu$NS cross section in the presence of a neutrino
magnetic moment interaction consists of the SM term in
Eq.(\ref{eq:CEvNS-xsec}) and a new term given by \cite{Vogel:1989iv}
\begin{equation}
  \label{eq:nmm_xsec}
  \frac{d\sigma_\gamma}{dE_r}=\pi\alpha^2_\text{em}\,Z^2
  \frac{\mu_\text{eff}^2}{m_e^2}
  \left(\frac{1}{E_r} -  \frac{1}{E_\nu}\right)F^2(q^2)\ ,
\end{equation}
where $\alpha_\text{em}$ is the fine structure constant and $\mu_\text{eff}^2$ is an effective parameter (in Bohr magneton
units $\mu_B$) that encodes the neutrino magnetic and electric dipole
moments (and transitions) along with neutrino oscillation
probabilities \cite{Borexino:2017fbd,AristizabalSierra:2020zod}. The
main feature of the new coupling is spectral distortions, resulting
from the fact that the cross section is forward peaked because of the
Coulomb divergence (infinite range interaction). This means that one
expects the WIMP discovery limit to be diminished at low WIMP masses
(low momentum transfer region). Moreover,  given the tight constraints on the
coupling implied by searches at GEMMA, BOREXINO and TEXONO
\cite{Beda:2010hk,Borexino:2017fbd,TEXONO:2009knm}, the discovery limit is also expected to converge to the
``standard'' case as soon as the momentum transfer reaches larger
values.

For the calculation of the impact of such interaction on WIMP
discovery limits we have fixed
$\mu_\text{eff}=2.9\times 10^{-11}\mu_B$, which corresponds to the
$90\%$CL reported by GEMMA \cite{Beda:2010hk} and XENON1T~\cite{XENON:2020rca}. The result is displayed
in Fig. \ref{fig:nmm}, which shows along with the discovery limits
implied by the new interaction the ``standard'' limits for
comparison. One can see that up to WIMP masses of order $\sim 0.2\,$GeV
the discovery limit worsens, because of the background enhancement
induced by the neutrino magnetic moment contribution. As soon as one
enters the region of large transfer momentum, the Coulomb divergence
fades away and the discovery limit matches that of the SM
alone. 
\section{Conclusions}
\label{sec:conclusions}
With the advent of the DM multi-ton detectors era in mind and with
well-established measurements of the CE$\nu$NS process by the COHERENT
collaboration, we have reconsidered the case of WIMP discovery
limits. We have adopted, for the first time, a data-driven analysis in which we have
treated the CE$\nu$NS cross section as a parameter entirely determined
by experimental data. Using this approach, while taking into account
neutrino flux uncertainties, we have derived WIMP discovery limits
using the CsI and LAr COHERENT data sets
\cite{Akimov:2017ade,COHERENT:2020iec}. Our results are free from
theoretical and phenomenological assumptions. They are also of particular
interest for future experiments XENONnT, DARWIN, DarkSide-20k and ARGO, as they fall in the region where
these experiments will have maximum sensitivities
\cite{Aprile:2015uzo,Aalbers:2016jon,DarkSide-20k:2017zyg,Billard:2021uyg,XENON:2020kmp}. 

We have as well presented a more assumption-dependent analysis in
which we have evaluated WIMP discovery limits by taken into account:
(i) Nuclear form factor uncertainties, (ii) possible fluctuations of
the weak mixing angle at low energies. Case (i) is driven by
uncertainties on the xenon point-neutron distribution
mean-square-radius. This quantity, in contrast to its proton
counterpart, has not been measured and so implies a sizable 
uncertainty on both the WIMP and CE$\nu$NS event rates. We have shown
that its effect on WIMP discovery limits, however, is mild and
relevant only in the 1-10 GeV WIMP mass region, where mainly $^8$B and
hep neutrino backgrounds matter. In case (ii), the effect is only
relevant for the CE$\nu$NS event rate and in regions of small WIMP
masses, where statistics is large enough to allow the proton
contribution to leave its imprint.

Additionally, we have presented a full model-dependent analysis aiming at
illustrating the impact that new physics at the neutrino background
level---lurking at low-scales---might have on WIMP discovery
limits. We considered light vector and scalar mediators as
well as neutrino magnetic moments/transitions. For a light vector
mediator, we have found that its presence can worsen WIMP discovery
limits by several orders of magnitude for WIMP masses up to $\sim10$ GeV. From that point on, our results indicate a crossover where the new
vector interaction actually leads to an improvement of the discovery
limit, albeit mild. WIMP discovery limits in the presence of scalar
and neutrino magnetic moment/transition interactions are always
worsen, in particular in regions of light WIMPs.

Finally, we point out that searches for WIMPs using forthcoming
multi-ton detector technologies require a precise understanding of
WIMP discovery limits. In our view, this calls---ideally---for the
most assumption-free determination of the effects of the neutrino
background, for the inclusion of known sub-leading effects and the
consideration of possible new physics effects. This has been the main
goal of the analyses presented here.

\section*{Acknowledgments}
We are grateful to Nicol\'as Rojas-Rojas for collaborating in the early stages of this project.
VDR acknowledges financial support by the SEJI/2020/016 grant (project ``Les Fosques'') funded by
Generalitat Valenciana, by the Universitat de Val\`encia through the sub-programme ``ATRACCI\'O DE TALENT 2019''
and by the Spanish grant ID2020-113775GB-I00 (AEI/10.13039/501100011033). 
The work of LJF was partially
supported by a postdoctoral CONACYT grant, CONACYT
CB2017-2018/A1-S-13051 (M\'exico) and DGAPA-PAPIIT
IN107118/IN107621.
The work of DKP is co-financed by Greece
and the European Union (European Social Fund- ESF) through the
Operational Programme ``Human Resources Development, Education and
Lifelong Learning'' in the context of the project ``Reinforcement of
Postdoctoral Researchers - 2nd Cycle" (MIS-5033021), implemented by
the State Scholarships Foundation (IKY). 
\\
\appendix
\section{Extraction of CE$\nu$NS cross section from COHERENT data}
\label{sec:extraction}
The datasets available for the CsI and LAr COHERENT detectors provide spectral information on the measured number of CE$\nu$NS events and their uncertainties~\cite{COHERENT:2018imc,COHERENT:2020ybo}. Under the assumption that the experimental cross section is proportional to its theoretical prediction, the COHERENT collaboration has already provided a determination of the measured CE$\nu$NS cross section on argon as $\sigma_\mathrm{meas} = \frac{N_\mathrm{meas}}{N_\mathrm{th}} \sigma_\mathrm{th}$ (for details see Ref.~\cite{Zettlemoyer:2020kgh}), where $N_\mathrm{meas}$ and $N_\mathrm{th}$ are the total number of measured and theoretical events, respectively. Motivated by the latter, here we perform a similar analysis which in our case is applied independently for each energy bin by considering $\sigma_\mathrm{meas}^i = n_\sigma^i \,\sigma_\mathrm{th}^i $. Following this approach, we extract the measured cross section along with its uncertainty for both CsI and LAr datasets. 

For the case of CsI we adopt the $\chi^2$ function
\begin{align}
	\chi^2_i &=   \left[\frac{N_\mathrm{exp}^i - (1+\alpha)N_\mathrm{meas}^i  (n^i_\sigma) - (1+\beta)B_\mathrm{0n}^i}{\sigma_\mathrm{stat}^i} \right]^2 \nonumber \\ 
	&+ \left(\frac{\alpha}{\sigma_\alpha}\right)^2 + \left(\frac{\beta}{\sigma_\beta}\right)^2 \;,
	\label{eq:chiSqFunCsI}
\end{align}
where $\alpha$ and $\beta$ are nuisance parameters which account for the uncertainty on the rate with $\sigma_\alpha=28\%$ and on the prompt neutron background $B_\mathrm{0n}$ with $\sigma_\beta=25\%$, respectively. Finally, the statistical uncertainty is defined as $\sigma_\mathrm{stat}^i = \sqrt{N_\mathrm{exp}^i + B_\mathrm{0n}^i +  2 B_\mathrm{ss}^i}$, where $B_\mathrm{ss}^i$ denotes the steady state background (for details see Ref.~\cite{COHERENT:2018imc}). 
For the case of LAr, we focus on the \emph{analysis-A}
of COHERENT~\cite{COHERENT:2020iec} and we follow the $\chi^2$ function~\cite{Cadeddu:2020lky}
\begin{equation*}
\begin{aligned}
\chi^2_i = & 
\dfrac{\left(
N^{i}_{\text{exp}}
-
\alpha N^i_{\mathrm{meas}}(n_\sigma^i)
-
\beta B^i_{\mathrm{PBRN}}
-
\gamma B^i_{\mathrm{LBRN}}\right)^2}
{\left( \sigma^i_{\mathrm{exp}} \right)^2 + \left[ \sigma_{\mathrm{BRNES}} \left( B^i_{\mathrm{PBRN}} + B^i_{\mathrm{LBRN}}\right)\right]^2}
\\ 
+&
\left( \dfrac{\alpha-1}{\sigma_\alpha} \right)^2
+
\left( \dfrac{\beta-1}{\sigma_\beta} \right)^2
+
\left( \dfrac{\gamma-1}{\sigma_\gamma} \right)^2 .
\label{chi-spectrum}
\end{aligned}
\end{equation*}
Here, BRNES corresponds to the Beam Related Neutron Energy Shape, while PBRN (LBRN) represents the Prompt (Late) Beam-Related Neutron Background data with $\sigma_\beta=32\%$ ($\sigma_\gamma=100\%$)~\cite{COHERENT:2020ybo}. The remaining parameters: Beam Related Neutron Energy Shape (BRNES) uncertainty $\sigma_\text{BRNES}=1.7\%$  and the systematic uncertainty  of the signal rate $\sigma_\alpha=13.4\%$ are taken from the estimations of Ref.~\cite{Cadeddu:2020lky}.
\\

\bibliographystyle{utphys}

\bibliography{references}
\end{document}